\documentclass[%
 reprint,
superscriptaddress,
 amsmath,
 amssymb,
 aps,
]{revtex4-1}

\usepackage{graphicx}% Include figure files
\usepackage{dcolumn}% Align table columns on decimal point
\usepackage{bm}% bold math
\usepackage[caption=false]{subfig}
\usepackage{mathtools}% http://ctan.org/pkg/mathtools
\usepackage{xcolor}

\usepackage{CJKutf8}

\renewcommand{\d}{\mathrm{d}}
\DeclareMathOperator{\sech}{sech}

\begin{document}

\begin{CJK*}{UTF8}{gbsn}

\preprint{APS/123-QED}

\title{Kink-antikink asymmetry and impurity interactions in topological mechanical chains}% Force line breaks with \\

% \thanks{A footnote to the article title}%

\author{Yujie Zhou (周宇杰)}
\author{Bryan Gin-ge Chen}
\altaffiliation[Current Address: ]{Department of Physics, University of Massachusetts, Amherst, Massachusetts, 01002, USA} 
\author{Nitin Upadhyaya}
\altaffiliation[Current Address: ]{Division of Physics and Applied Physics, School of Physical and Mathematical Sciences, Nanyang Technological University, Singapore 637371, Singapore} 
\author{Vincenzo Vitelli}
\affiliation{Instituut-Lorentz, Universiteit Leiden, 2300 RA Leiden, The Netherlands}

\date{\today}% It is always \today, today,
             %  but any date may be explicitly specified

\begin{abstract}

We study the dynamical response of a diatomic periodic chain of rotors
coupled by springs, whose unit cell breaks spatial inversion symmetry.
In the continuum description, we derive a nonlinear field theory which
admits topological kinks and antikinks as nonlinear excitations but
where a topological boundary term breaks the symmetry between the two
and energetically favors the kink configuration. Using a cobweb plot,
we develop a fixed-point analysis for the kink motion and demonstrate
that kinks propagate without the Peierls-Nabarro potential energy
barrier typically associated with lattice models. Using continuum
elasticity theory, we trace the absence of the Peierls-Nabarro barrier
for the kink motion to the topological boundary term which ensures
that only the kink configuration, and not the antikink, costs zero
potential energy. Further, we study the eigenmodes around the kink and
antikink configurations using a tangent stiffness matrix approach
appropriate for pre-stressed structures to explicitly show how the
usual energy degeneracy between the two no longer holds. We show how
the kink-antikink asymmetry also manifests in the way these nonlinear
excitations interact with impurities introduced in the chain as
disorder in the spring stiffness. Finally, we discuss the effect of
impurities in the (bond) spring length and build prototypes based on
simple linkages that verify our predictions.
 
%\begin{description}

  % \item[Usage]
% Secondary publications and information retrieval purposes.
% \item[PACS numbers]
% May be entered using the \verb+\pacs{#1}+ command.
% \item[Structure]
% You may use the \texttt{description} environment to structure your abstract;
% use the optional argument of the \verb+\item+ command to give the category of each item. 
% \end{description}

\end{abstract}

% \pacs{Valid PACS appear here}% PACS, the Physics and Astronomy
%                              % Classification Scheme.
% %\keywords{Suggested keywords}%Use showkeys class option if keyword
%                               %display desired

\maketitle

%\tableofcontents

\section{Introduction}

Topological ideas have led to recent advances in continuum mechanics often inspired by the
physics of electronic topological insulators and the quantum Hall effect. In these electronic
systems the basic question is whether a material is an insulator or a conductor. The
answer depends on which portion of a topological insulator one examines: the bulk is usually gapped and
hence insulating while the edge displays gapless edge modes whose existence is protected
from disorder and variations in material parameters by the existence of integer-valued
topological invariants \cite{Hasan2010}. In topological mechanical systems, the corresponding question is
whether a material is rigid or floppy. The ability to modulate the rigidity of a structure
in space allows to robustly localize the propagation of sound waves
\cite{Prodan2009,Berg2011,Susstrunk2015,Susstrunk2016,Wang2015a,Wang2015b,Mousavi2015,Khanikaev2015,Nash2015,Yang2015,Yang2015a,Xiao2015a,Peano2015,Kariyado2015,Deymier2015,Bi2015,Po2014,Lawler2015},
change shape in selected portions
\cite{Kane2013a,Lubensky2015,Vitelli2012a,Chen2014,Vitelli2014,Meeussen2016,Paulose2015,Chen2016,Rocklin2015,Rocklin2016}
or focus stress leading to selective buckling or failure \cite{Paulose2015a}.

By translating the topological properties of bands of electronic states into the classical
setting of vibrational bands, one can identify topologically protected and hence robust
properties of vibrational modes in both discrete lattices and continuous media. For example, the concept of 
``topological polarization'' recently introduced by Kane and Lubensky \cite{Kane2013a} building on counting ideas from  
Maxwell and Calladine ~\cite{Maxwell1864,Calladine1978} determines the
existence and the position of zero-energy motions that are localized at edges
and defects of a marginally rigid mechanical lattice (one in which constraints and degrees of freedom are exactly balanced).

\begin{figure}[h!]
  \subfloat[]{\label{fig:configuration}  
    \includegraphics[width=0.45\textwidth]{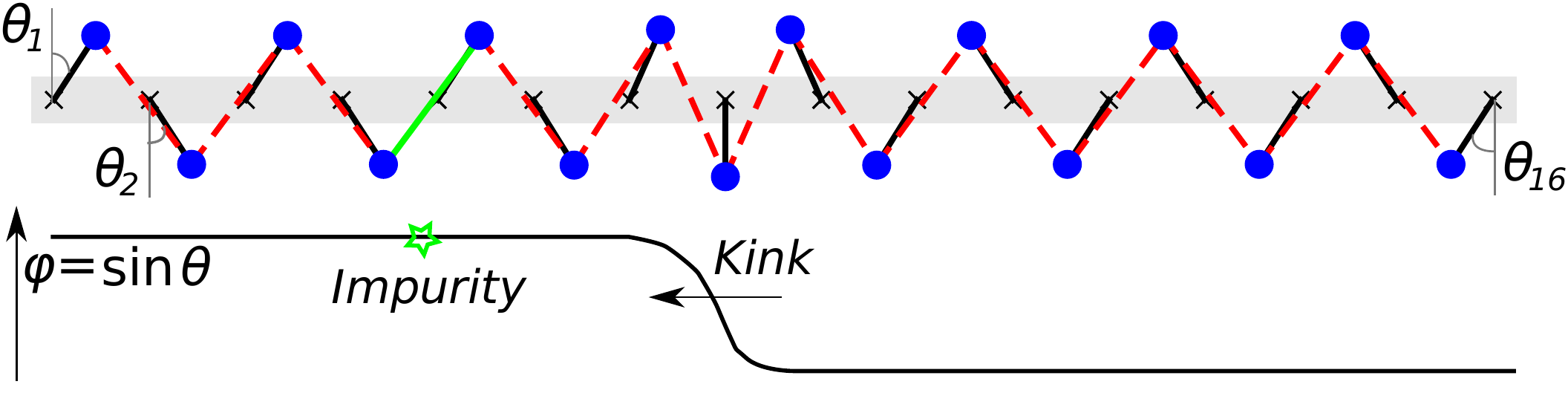}}
  
  \subfloat[]{
    \includegraphics[width=0.45\textwidth]{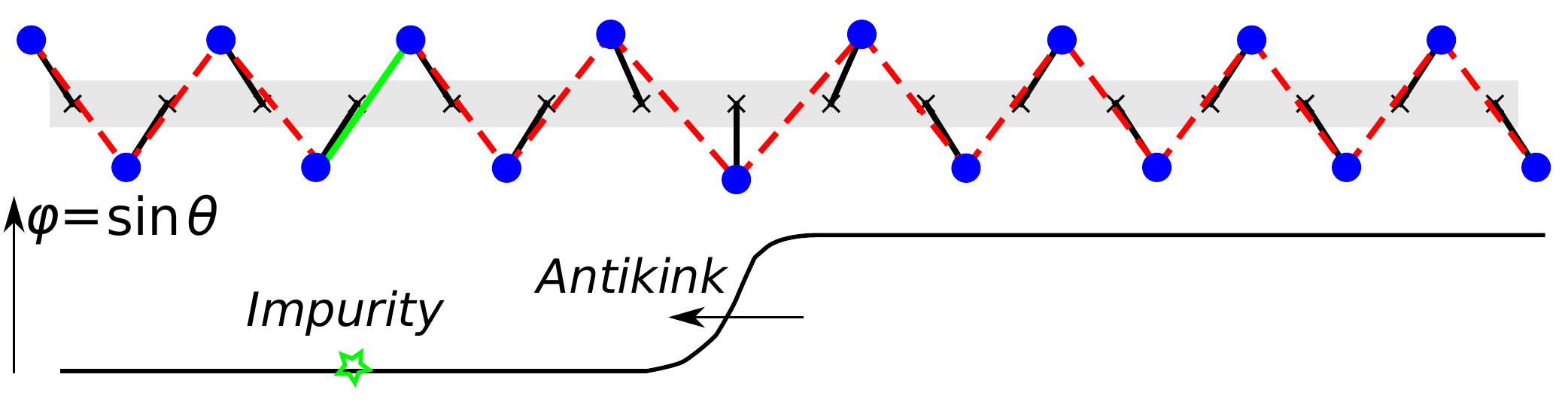}}

    \subfloat[]{  \label{fig:two-rotor}
  \includegraphics[width=0.25\textwidth]{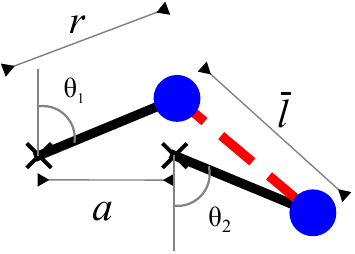}}

\caption{A kink~(\textbf{a}) and an antikink~(\textbf{b}) configuration in a topological
  chain~(TC) model of rotors~(blue) and springs~(red dashed lines) in the presence of a
  single impurity (green solid lines) modeled as a spring with a different stiffness. For
  the kink profile, the springs in the chain are at their rest length, while for the
  antikink, they are stretched. A sketch of kink and antikink profiles in terms of the
  continuum field variable $u=\sin\theta$ (where $\theta$ is the rotor angle) is shown
  below each configuration. (\textbf{c}) A two-rotor system. The masses are the blue dots,
  the rigid rotors are the black lines, the pivots are the crosses, and the spring is the
  dashed red line. Here, $a$ is the lattice spacing, $r$ is the rotor length, $\bar{l}$ is
  the rest length of the springs and $\theta _{1,2}$ are the rotor angles with respect to
  the vertical.} 
\end{figure}

Perhaps the simplest model of topological mechanical lattices is the rotor chain proposed
in Ref.~\cite{Kane2013a}. The system consists of a chain of classical rotors harmonically
coupled with their nearest neighbours, as shown in Fig.~\ref{fig:configuration}. There are
two distinct classes of ground state configurations, one with all rotors leaning towards
the left and the other where they lean towards the right. Mathematically, these two states
may not be deformed to each other without the appearance of bulk zero modes; thus they may
each be assigned a different winding number, associated with the Fourier transform of the
compatibility matrix $C(q)$, which connects the linear displacement of rotors with the
extension of bonds; see Ref.~\cite{Lubensky2015} for a detailed explanation.

The above considerations arise from band theory and thus concern only the linearized
zero-energy infinitesimal motions. Indeed, the vanishing of the linear response implies
that nonlinear effects dominate. By developing a nonlinear theory of the rotor chain, it
was shown in Ref. \cite{Chen2014} that
the infinitesimal zero-mode displacement integrates to a finite motion. This motion can be described in the continuum limit
by objects similar to ``kinks'' in the $\phi^4$ field theory ~\cite{Manton}, which
connects the topological polarization invariant of the linear vibrations to the study of
topological solitons ~\cite{Chen2014,Vitelli2014}. Although the two appearances of the
term ``topology'' in the linear and nonlinear theory stem from different contexts, the
latter encompasses the predictions of the former and also explains additional features
exclusive to the nonlinear dynamics
~\cite{Vitelli2014}.

The nonlinear dynamics of this topological chain can be approximated by the critical
trajectories of a Lagrangian written in the following form ~\cite{Chen2014,Vitelli2014}
\begin{align}
\label{eq:introlagrangian}
\begin{split}
  L&= \int \d x \Big(\frac{\partial u}{\partial t}\Big)^2- \Big(\frac{\partial
      u}{\partial x}\Big)^2 \\
    & ~~~~~~~~~~~~~~~~~~ -\frac{1}{2}(u^2-1)^2-\sqrt{2}~\frac{\partial u}{\partial x}(u^2-1) .
\end{split}  
\end{align}

The first term corresponds to the kinetic energy while the second and third are the ones encountered for example in the Landau theory of the Ising model. 
Note, however, that there is an additional boundary term that contributes to the energy
but does not enter the Euler-Lagrange equation. Hence, one obtains static kink and
antikink solitary wave solutions of the usual form ~\cite{Manton}

\begin{equation}
  u=\pm\tanh\Bigg(\frac{x-x_0}{\sqrt{2}}\Bigg).
\end{equation}
The boundary term gives new properties to the solutions and breaks the
symmetry between kinks and antikinks. For example, it predicts that
the static kink configuration costs zero potential energy while the
static antikink configuration has a finite potential energy.  Previous
work on this model has been motivated by the kink's zero-energy properties, and thus
the shape and stability of the antikink and its dynamical behavior were not studied.

\begin{figure}[h!]
  \includegraphics[width=0.45\textwidth]{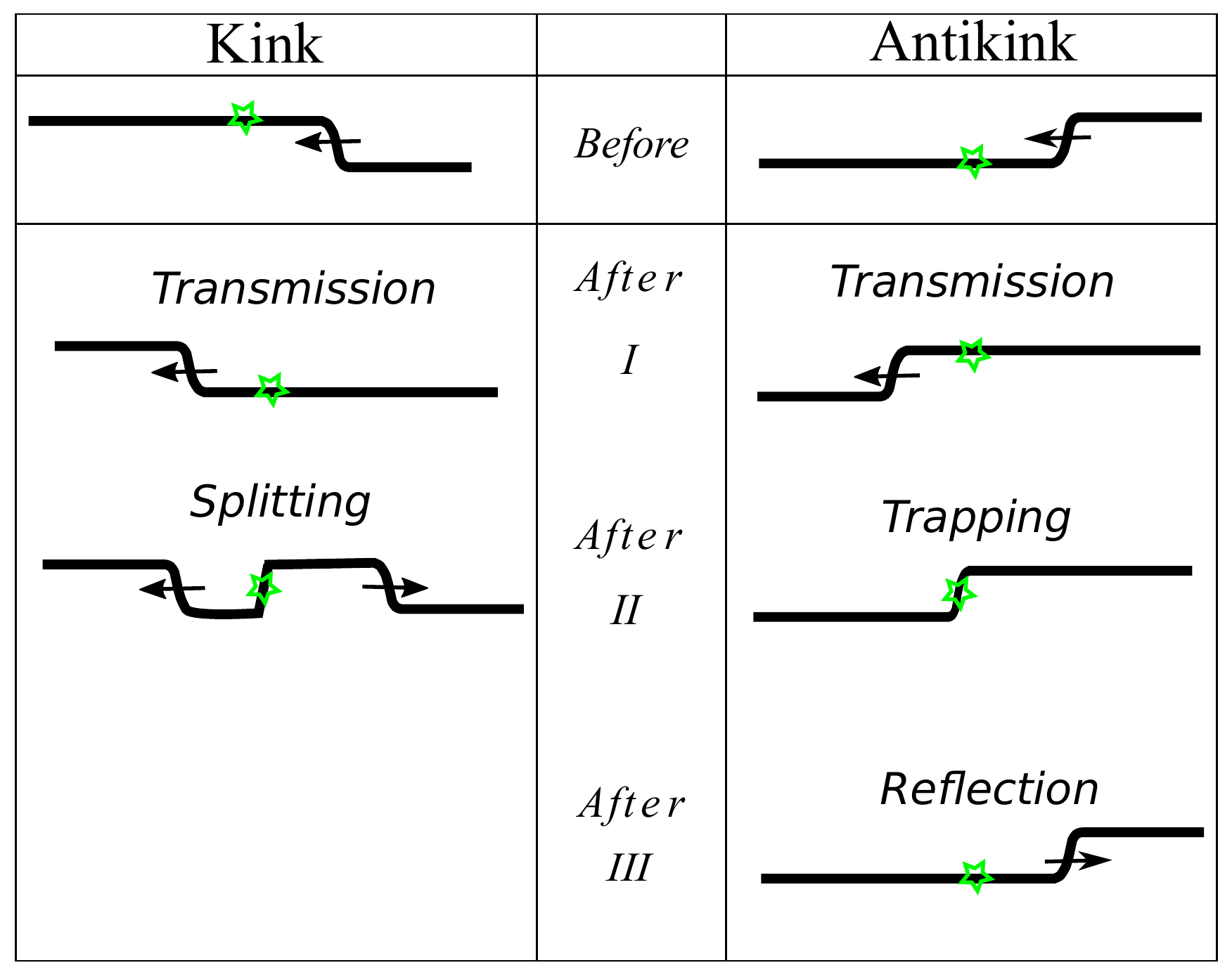}
\captionsetup{justification=centerlast,margin=0cm,singlelinecheck=false}
\caption{Illustrated are the possible scenarios for how the kink and antikink interact
  with a single impurity of spring stiffness. As indicated by the arrow, an initial kink
  or antikink approaches the impurity site (indicated by the green star) from the right.
  After scattering, the incident kink is either: (\textbf{I}) perfectly transmitted or
  (\textbf{II}) splits into a reflected kink, a transmitted kink and an antikink that gets
  trapped at the impurity site. The incident antikink is either: (\textbf{I}) perfectly
  transmitted, (\textbf{II}) trapped at the impurity site or (\textbf{III}) perfectly
  reflected.}
\label{fig:scheme-scattering}
\end{figure}

In this paper we explore the physics of these finite-energy configurations. We compare the
dynamics of the kink and antikink sectors in the topological rotor chain and study their
interaction with a lattice impurity. We find that differences arising from the topological
boundary term are apparent in all of these aspects. In Section II, we explain the discrete
model and develop a fixed-point analysis of the kink motion using a cobweb plot. In
Section III, we review the continuum theory and compare the predictions for the antikink
with the discrete model. In Section IV, we study the eigenmodes of the chain around a
single kink or antikink profile. We exploit the tangent stiffness matrix approach
developed by Guest \cite{Guest2006} to analyze prestressed structures. In Section V, we
study the nonlinear transport properties. In a conventional continuum $\phi^4$ field
theory, owing to translation invariance, both the kink and antikink propagate at uniform
speed. However, lattice discreteness effects breaks this invariance and generates the
so-called Peierls-Nabarro (PN) barrier~\cite{Combs1983,Braun1998,Roy2007}. For the
topological rotor model, we find that only the antikink has a finite PN barrier whereas
the kink always propagates freely. We explain this phenomenon as a consequence of the
zero-energy cost associated with the kink profile. In Section VI, we investigate how kinks
and antikinks interact with a spring constant impurity in the lattice. For the normal
$\phi^4$ model, a phenomenological theory predicts alternating windows of initial kink
(antikink) velocities that leads to reflection, trapping and transmission of the
excitation~\cite{Fraggis1989, Fei1992}. By contrast, for the topological rotor model that
we study, an impurity in the spring stiffnesses results in dramatically different
scattering behaviors for the kink and antikink respectively.
Fig.~\ref{fig:scheme-scattering} summarizes all the possible scattering scenarios that we
observe. Finally, in Section VII, we make a connection between linear mode analysis and
nonlinear dynamics of kink motion in the context of spring length impurities. We conclude by
listing some open questions related to our study.

\section{Discrete model and cobweb plot}\label{sec:cobweb}

The model we study consists of rotors of length $r$. The rotor pivots are placed on a 1D lattice with spacing $a$. The angles $\theta_i$ of the rotors are measured in an alternating
fashion along the lattice, from the positive $y$-axis at odd-numbered sites and negative $y$-axis at even-numbered sites. The equilibrium angle is $\overline{\theta}$ for a uniform lattice configuration \textit{without} a kink or antikink. The masses $M$ at the tips of the rotors are connected by harmonic springs with identical rest lengths $\overline{l}$ and
spring constants $k$. The two-rotor unit cell of the topological chain is illustrated in Fig.~\ref{fig:two-rotor}.

We now construct the chain \textit{with} a kink under free boundary conditions. There are $n$ rotors and $n-1$ springs. If we assume that the springs are infinitely stiff~($k\rightarrow\infty$), the springs become $n-1$ constraints and the system only has a single independent degree of freedom. The angle of a single rotor determines all the others iteratively. This degree of freedom manifests itself as a mechanism which, as has been previously shown in~\cite{Chen2014}, can be approximately described by the domain wall solution in a modified $\phi^4$ theory \footnote{Varying the parameters~($a,r,\overline{\theta}$) yield other phases of the topological rotor chain. In this work, we only consider the topological chain in the \textit{flipper} phase~\cite{Chen2014} where the $\phi^4$ theory is a valid approximation. The name flipper describes the back-and-forth motion of the rotors as a kink propagates, in contrast to the \textit{spinner} phase, where the rotors complete a full circle. The continuum limit of the spinner phase can be approximately described by the sine-Gordon theory}. 
We call this mechanism a ``kink'' and discuss its continuum theory in the following sections.

\begin{figure}[h!]
  \subfloat[]{\label{fig:kconfig}
    \includegraphics[width=0.45\textwidth]{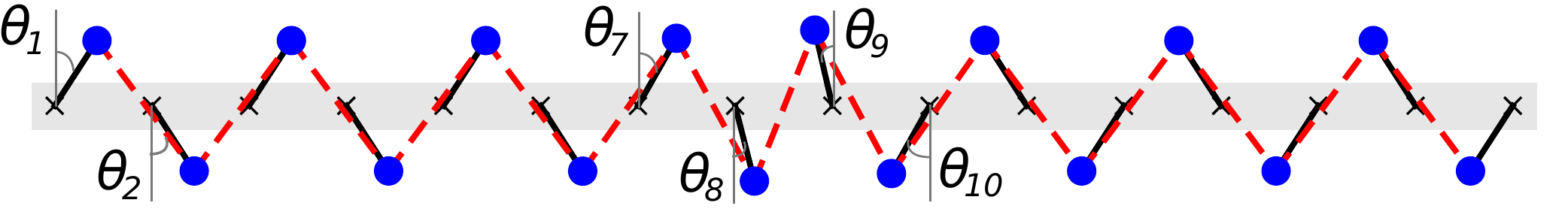}}
  
  \subfloat[]{\label{fig:kcobwebplot}
  \includegraphics[width=0.45\textwidth]{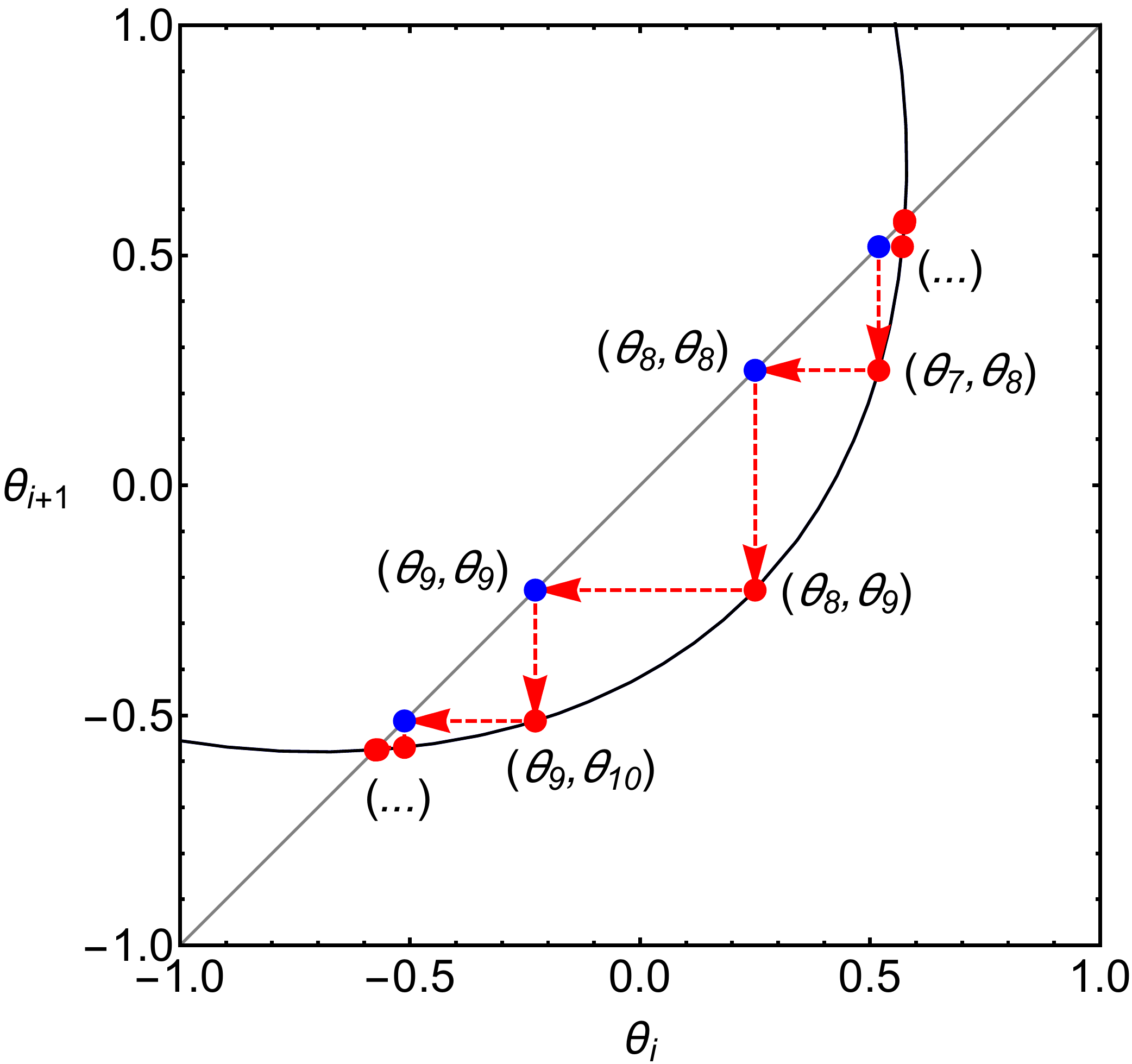}}
\caption{\label{fig:kcobweb}The configuration~(\textbf{a}) and the corresponding cobweb plot~(\textbf{b}) for the kink in a topological rotor chain with $r/a=0.8$, $|\overline{\theta}|=0.58$. The springs are at their rest lengths. In (\textbf{b}), the black curve is the constraint equation which ensures that the  springs are unstretched, the gray diagonal line satisfies $\theta _{i+1}=\theta _i$, the blue point $(\theta_{i},\theta_i)$ represents rotor $i$, the red point $(\theta_i,\theta_{i+1})$ represents the spring connecting rotors $i$ and $i+1$, and the red
dashed lines with arrows indicates the iterative process that generates the kink profile. The iteration steps from $\theta_7$ to $\theta_{10}$ are shown.} 
\end{figure}

We use a cobweb plot to display the kink in Fig.~\ref{fig:kcobweb}.
This is a tool for
visualizing the process of iteratively solving the nonlinear constraint equations
Eqn.~(\ref{eq:spring}) cell by cell. We construct the cobweb plot by drawing
(1) a diagonal line $\theta_i = \theta_{i+1}$ and (2) a curve of the implicit function given
by the nonlinear constraint equation that ensures the springs are not stretched,
\begin{equation}
  \label{eq:spring}
(a+r\sin\theta_i-r\sin\theta_{i+1})^2+(r\cos\theta_i+r\cos\theta_{i+1})^2=\overline{l}^2.
\end{equation}
(An explicit relation between neighbouring rotor angles is derived analytically
with complex notation in Appendix \ref{App: AppendixA}.)

~

~
The iteration steps are as follows:
\begin{enumerate}
\item Given the angle $\theta_1$ of the first rotor at the left end, find the point on the
  function curve with coordinates $(\theta_1,\theta_2)$.
\item Draw a horizontal line from $(\theta_1,\theta_2)$ to the diagonal line.
  This gives the point $(\theta_2,\theta_2)$.
\item Draw a vertical line from $(\theta_2,\theta_2)$ to the function curve. This gives
  the point $(\theta_2,\theta_3)$.
\item Repeat step 2 and 3 until the point $(\theta_{n-1},\theta_n)$ is found.
\end{enumerate}

In Fig.~\ref{fig:kcobwebplot}, we illustrate steps 2 and 3 from $\theta_7$ to $\theta_{10}$, which are near the kink center. The blue point with coordinates $(\theta_{i},\theta_i)$ stands for the $i$th rotor of angle $\theta_{i}$. The red point with coordinates $(\theta_{i},\theta_{i+1})$ represents the state of the spring that connects the rotors of $\theta_{i}$ and $\theta_{i+1}$.

Note that in Fig.~\ref{fig:kcobwebplot}, the diagonal line and the function curve intersect at two points. They are the fixed points of iteration. If all the red points $(\theta_i,\theta_{i+1})$ stay at one fixed point, the plot represents a uniform lattice. The iteration step proceeds from the leftmost rotor of the chain to the rightmost. We see that the flow proceeds outwards from one fixed point and then inwards towards  the other fixed point.  

The cobweb plot may be used to graphically derive the decay lengths of
zero energy deformations, as they approach their uniform limits. As
mentioned above, a fixed point corresponds to an intersection between
the line $\theta_{i}=\theta_{i+1}$ and the function curve. Note that the behavior of $\theta_i$ as it approaches a fixed point resembles a "self-similar" zigzag motion between $\theta_i=\theta_{i+1}$ and the tangent line of the function curve.  This motivates linearizing the function curve around the fixed point as follows:
\begin{equation}
\theta_{i+1}-\overline{\theta}=F'(\overline{\theta})(\theta_i-\overline{\theta}),
\end{equation}

\noindent where $\overline{\theta}$, the equilibrium angle, is also just the value of the fixed-point angle and $F'(\overline{\theta})$ is the slope of the function curve at that point (which could be computed explicitly in terms of $r,a,\overline{l}$). This equation yields that $\theta_i-\overline{\theta}\propto\exp(\log F'(\overline{\theta})i)$, or that the decay length is $|1/\log F'(\overline{\theta})|$ (the sign of $\log F'$ tells us whether the fixed point is attracting or repelling).  This result recovers the penetration depth of the boundary modes computed in Ref.~\cite{Chen2014} using band theory.

In the cobweb plot, the static kink appears as a sequence of points on the function curve interpolating between a repelling and attracting fixed point. The dynamics of the kink in the
cobweb plot is therefore the flow of a cascade of points between a pair of fixed points (Movie S1). While the kink propagates, the points in the middle, such as $(\theta_{7},\theta_{8})$, $(\theta_{8},\theta_{9})$ and $(\theta_{9},\theta_{10})$, corresponding to the kink center, move more than those points close to the fixed points, corresponding to the spatially localized nature of the kinetic energy.

\begin{figure}[h!]
  \subfloat[]{\label{fig:akconfig}
    \includegraphics[width=0.45\textwidth]{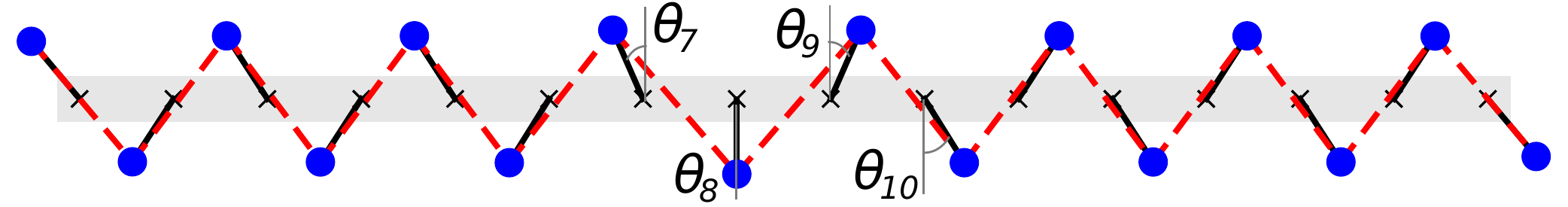}}
  
  \subfloat[]{\label{fig:akcobwebplot}
  \includegraphics[width=0.45\textwidth]{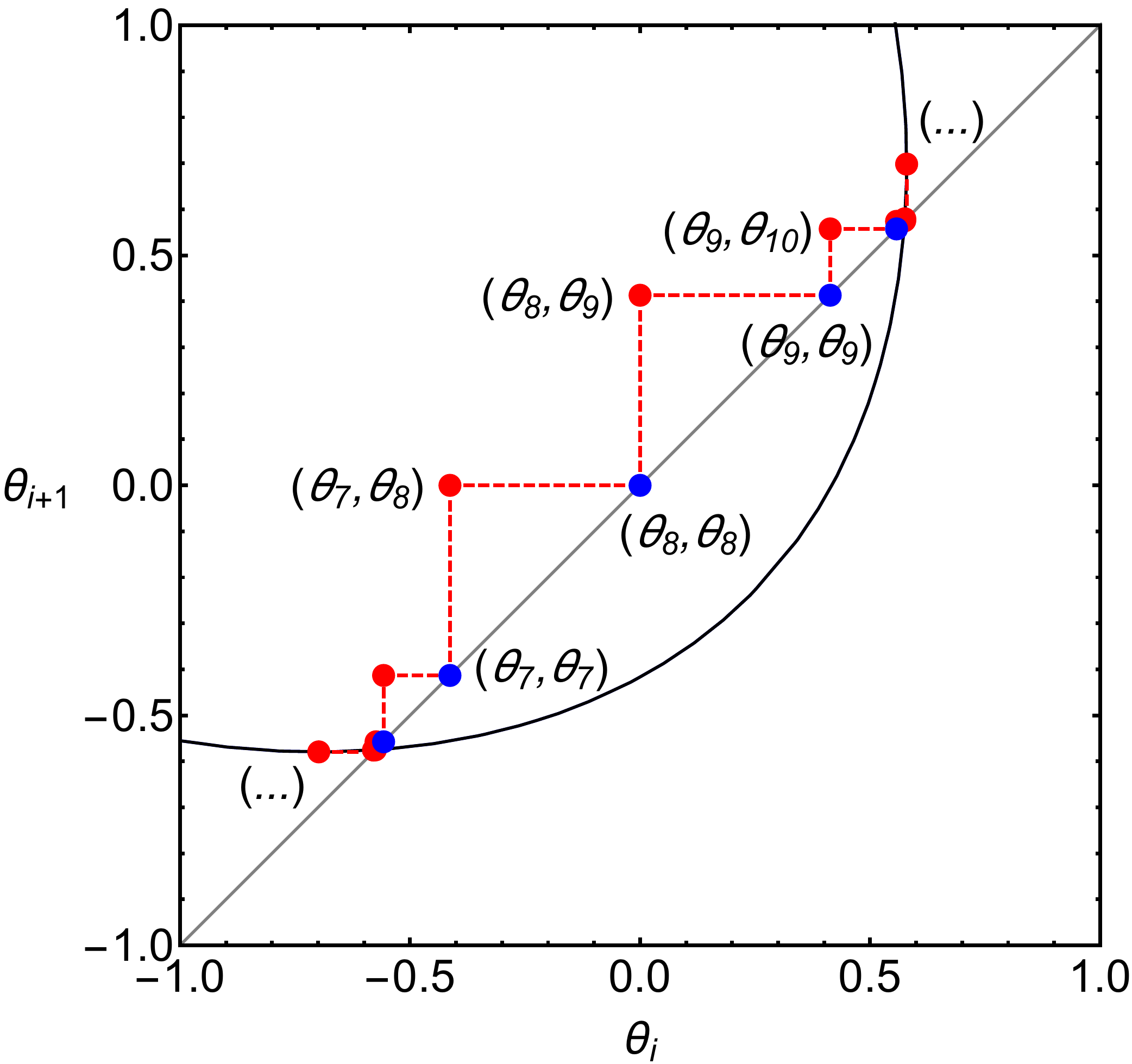}}
  \caption{\label{fig:akcobweb}The configuration~(\textbf{a}) and the corresponding cobweb plot~(\textbf{b}) for an antikink profile in the topological rotor chain with $r/a=0.8$, $|\overline{\theta}|=0.58$, where we see that the springs are stretched. In~(\textbf{b}) the same graphic notation as in Fig.~\ref{fig:kcobweb} is used except that we have not used an iterative process for constructing the antikink profile, rather, depicted is only a visualization of the configuration of the rotor chain. The red points are obtained by first reflecting the red points in Fig.~\ref{fig:kcobwebplot} across the diagonal line, and then relaxing the springs using dissipative Newtonian dynamics. Note that the two rotors at the edges need to be collinear with the springs to ensure force balance. This results in the angles overshooting at the fixed points.}
\end{figure}

\begin{figure}[h!]
  \subfloat[]{\label{fig:akprofile}
    \includegraphics[width=0.45\textwidth]{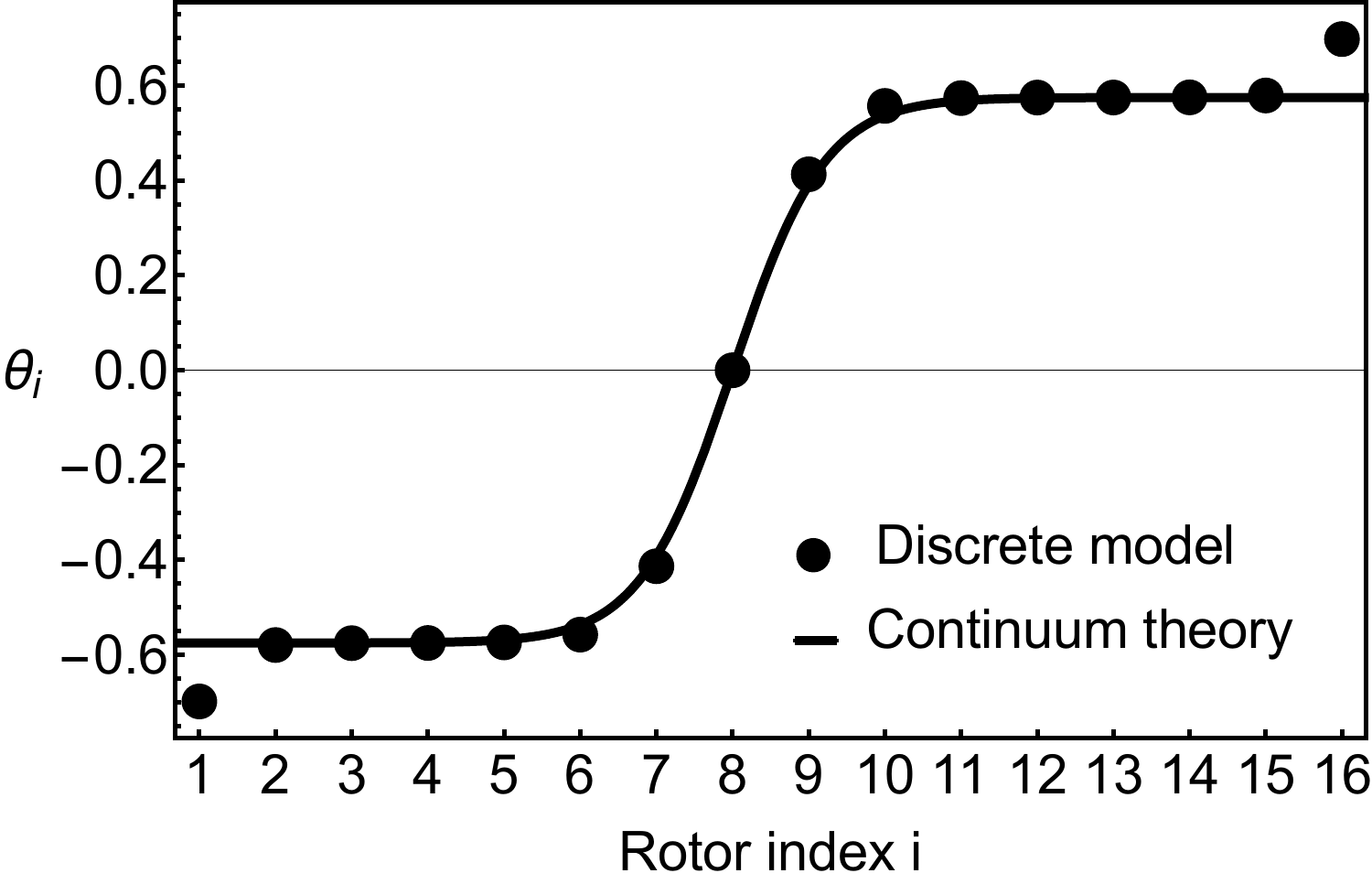}}
  
  \subfloat[]{\label{fig:akspringstretch}
  \includegraphics[width=0.45\textwidth]{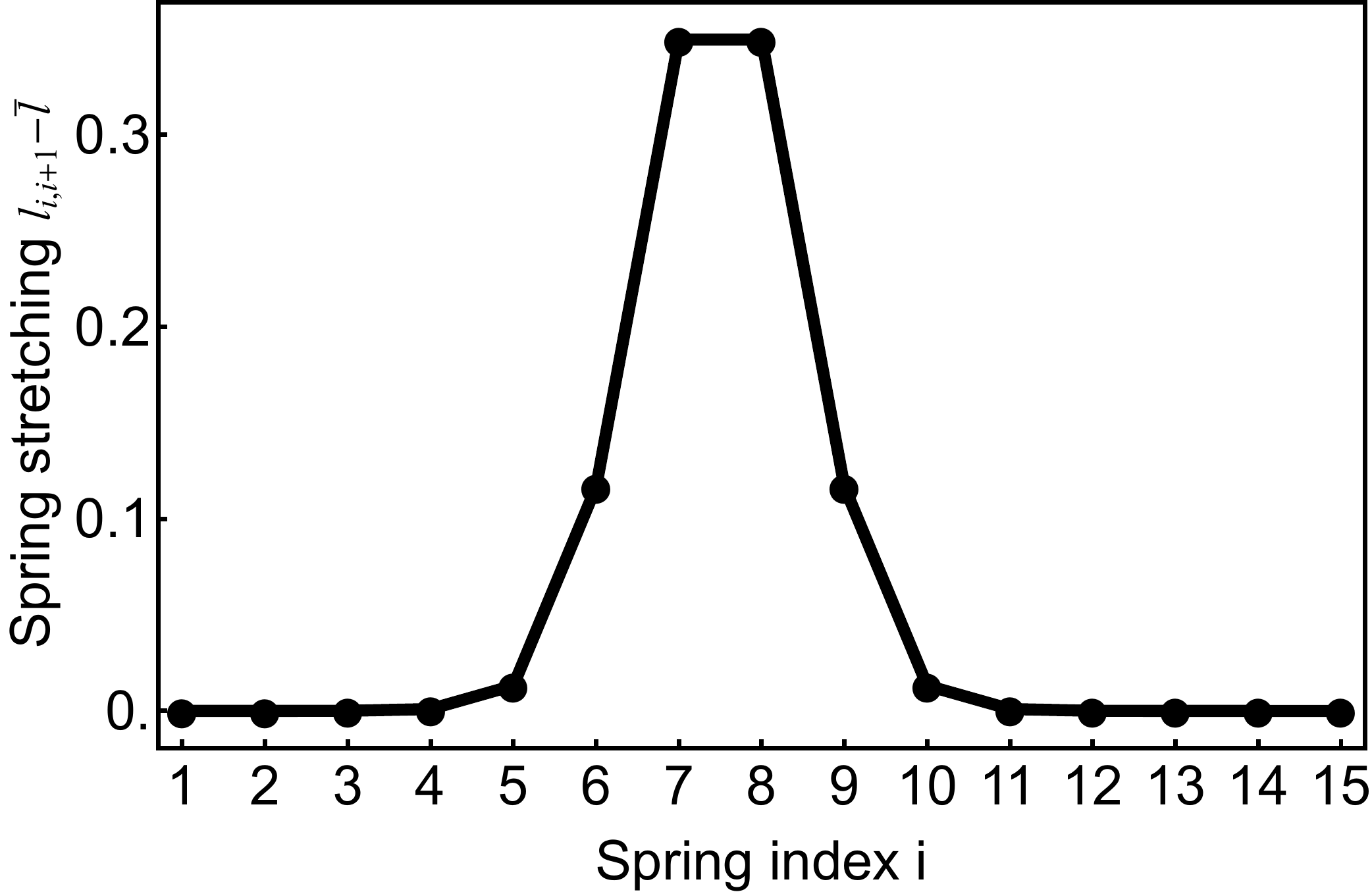}}
  \caption{\label{fig:aktheta}(\textbf{a}) The $\theta$ profile (rotor angles) for the antikink profile in Fig.~\ref{fig:akconfig} and the corresponding continuum prediction from Eqn.~(\ref{eq:kinksolution}). Note that the two rotors at the edges need to be collinear with the springs to ensure force balance and this results in the rotor angles overshooting the equilibrium value $\overline{\theta}=\pm0.58$. (\textbf{b}) The amount of spring stretching for the antikink profile.}
\end{figure}

Generating an antikink requires a few more steps, as it stretches
springs, and thus does not satisfy a constraint function that we could
iteratively solve. However, the continuum theory suggests that kinks
and antikinks both have the same functional profiles with only their
signs reversed (see Section III). As a result, we use the same
iterative procedure as that for the kink, and then simply swap the
appearances of $\theta_i$ and $\theta_{i+1}$ in Eqn.~(\ref{eq:spring}) to obtain an approximation for the antikink profile. This method is
equivalent to reflecting the red points in Fig.~\ref{fig:kcobwebplot} across the diagonal line. The antikink constructed this way is not an equilibrium configuration and has unbalanced stresses in the springs. This is because generically, the profiles of the kink and antikink are not the same in a discrete topological rotor chain.  We next relax the springs using
\textit{dissipative} Newtonian dynamics to remove the unbalanced stresses and obtain a
stable profile, which we show in the cobweb plot in
Fig.~\ref{fig:akcobweb}. In that figure, the spring connections (red dots) around the core
of the antikink profile (rotors 8 and 9) do not fall on the curve which corresponds to
unstretched springs. This implies large spring deformations which we show explicitly in
Fig.~\ref{fig:akspringstretch}. The amount by which the springs are stretched is
symmetrical around the 8th spring, which is in accordance with the fact that a stable
antikink has balanced forces on each rotor. Note that we have fixed the boundary
conditions to ensure that the antikink is in mechanical equilibrium, which is not
generically true. As discussed later in Section V, this has important consequences for the PN barrier.

\section{Continuum theory}

In this section, we review the continuum approximation to the kink and antikink profiles \cite{Chen2014} and compare these with the discrete model developed in the previous section. The discrete Lagrangian for the topological rotor chain (see also Fig.~\ref{fig:kconfig}) with free boundary conditions is
\begin{equation}
  \label{eq:discretelagrangian}
  L=\sum_{i=1}^{n}\frac{1}{2} M r^2\bigg(\frac{\d \theta_i}{\d t}\bigg)^2-\sum_{i=1}^{n-1}\frac{1}{2}k(l_{i,i+1}-\overline{l})^2.
\end{equation}
Here, $n$ is the total number of rotors, $M$ is the mass at the tip of a rotor, $r$ is the
rotor length, $\theta_i$ is the angle that rotor $i$ makes with the vertical (measured alternately as shown in Fig.~\ref{fig:kconfig}), $k$ is the spring constant, $\overline{l}$ is the rest length of the spring and $l_{i,i+1}$ is the instantaneous length of the spring that connects rotor $i$ to rotor $i+1$. From geometry 
\begin{align}
  \label{eq:instantlength}
\begin{split}
  l_{i,i+1}^2% &=(a+r\sin\theta_{i+1}-r\sin\theta_{i})^2+\\
  % &~~~~(r\cos\theta_{i}+r\cos\theta_{i+1})^2\\
  &=a^2+2ar(\sin\theta_{i+1}-\sin\theta_{i})+2r^2+\\
  &~~~~2r^2\cos(\theta_{i}+\theta_{i+1}).
\end{split}  
\end{align}
which in the uniform limit $\theta _i=\theta _{i+1}=\bar{\theta}$ gives the rest length of the spring $\overline{l}^2=a^2+4r^2\cos^2\overline{\theta}$.

We make the working assumption that deformations do not stretch the
springs significantly and hence we can neglect (or add) terms higher than
quadratic order in $l_{i,i+1}-\overline{l}$ for all $i$.  This is a 
reasonable approximation for the system configuration with a kink profile but is not well-justified for an antikink profile. However, in the limit that $\overline{\theta} \ll 1$, we find this to be a good approximation for both kinks and antikinks. Within this limit, we therefore express the potential energy term in Eqn.~(\ref{eq:discretelagrangian}) as
\begin{align}
  \label{eq:discretepotential}
\begin{split}
  \frac{1}{2}k(l_{i,i+1}-\overline{l})^2% &=\frac{1}{2}\frac{k}{(l_{i,i+1}+\overline{l})^2}(l_{i,i+1}+\overline{l})^2(l_{i,i+1}-\overline{l})^2  \\
                                        &\approx\frac{k}{8\overline{l}^2}\bigg(l^2_{i,i+1}-\overline{l}^2\bigg)^2.
\end{split}                                        
\end{align}
Substituting the expression for $\bar{l}$ and Eqn.~(\ref{eq:instantlength}) into Eqn.~(\ref{eq:discretepotential}), we express the potential energy as
\begin{align}
  \label{eq:discretepotential2}
\begin{split}
  V_{i,i+1}% &\approx\frac{k}{8\overline{l}^2}\big(l^2_{i,i+1}-\overline{l}^2\big)\\
  % &= \frac{k}{8\overline{l}^2} (2r^2)^2
  % \bigg(\frac{a}{r}(\sin\theta_{i+1}-\sin\theta_{i})+1\\
  % &+\cos(\theta_{i}+\theta_{i+1})-2\cos^2\overline{\theta}\bigg)^2\\
  &=\frac{kr^4}{2\overline{l}^2}    \bigg(\frac{a}{r}(\sin\theta_{i+1}-\sin\theta_{i})\\
  &~~~~-\cos2\overline{\theta}+\cos(\theta_{i}+\theta_{i+1}) \bigg)^2.
\end{split}                                        
\end{align}

Now we take the continuum limit of the potential. First we define a continuum field for
the rotor angles $\theta(x)$, where the spatial variable $x=ia+\frac{a}{2}$ is located
symmetrically between two rotors in the unit cell. To leading order,
$\theta_i\rightarrow\theta(x)-(a/2)(\d \theta/\d x)$ and
$\theta_{i+1}\rightarrow\theta(x)+(a/2)(\d \theta/\d x)$. Eqn.~(\ref{eq:discretepotential2})
can then be expressed as
\begin{align}
  \label{eq:continuouspotential}
\begin{split}
a V[\theta]% &=\frac{kr^4}{2\overline{l}^2} \bigg(\frac{a}{r}\Big(\sin\big(\theta+ \frac{a}{2} \frac{\d \theta}{\d x}\big)-\sin\big(\theta- \frac{a}{2} \frac{\d \theta}{\d x}\big)\Big)\\
% &~~~~~~~~~~~~~~~~~~~-\cos2\overline{\theta}+\cos2\theta \bigg)^2\\
% &=\frac{kr^4}{2\overline{l}^2} \bigg(\frac{a}{r}\Big(\sin\theta+\cos\theta \frac{a}{2}\frac{\d \theta}{\d x}-\big(\sin\theta-\cos\theta\frac{a}{2}\frac{\d \theta}{\d x}\big)\Big)\\
% &~~~~~~~~~~~~~~~~~~~-1+2\sin^2\overline{\theta}+1-2\sin^2\theta \bigg)^2\\
% &=\frac{kr^4}{2\overline{l}^2} \bigg(\frac{2a}{r}\big(\cos\theta \frac{a}{2}\frac{\d \theta}{\d x} \big)+2\sin^2\overline{\theta}-2\sin^2\theta\bigg)^2\\
% &=\frac{2k}{\overline{l}^2} \bigg(\frac{a^2r}{2}\cos\theta
% \frac{\d \theta}{\d x}+r^2\sin^2\overline{\theta}-r^2\sin^2\theta\bigg)^2\\
&=\frac{2k}{\overline{l}^2}\left(\frac{a^2}{2}\frac{\d u}{\d x}+\overline{u}^2-u^2\right)^2,
\end{split}                                        
\end{align}
where we have defined the projection of the rotor position on the $x-$axis as a new field
variable $u(x)\equiv r\sin\theta(x)$ and $\overline{u}\equiv r\sin\overline{\theta}$.

The kinetic energy density term in Eqn.~(\ref{eq:discretelagrangian}) then assumes the
form
\begin{align}
  \label{eq:continuouskinetic}
\begin{split}
  a T[\dot{\theta}]% =\frac{1}{2} M r^2\bigg(\frac{\d \theta}{\d t}\bigg)^2
  % &=\frac{1}{2}\frac{Mr^2}{r^2\cos^2\theta}\bigg(\frac{\d u}{\d t}\bigg)^2\\
  % &=\frac{1}{2}\frac{Mr^2}{r^2-r^2\sin^2\theta}\bigg(\frac{\d u}{\d t}\bigg)^2\\
  &=\frac{1}{2}\frac{Mr^2}{r^2-u^2}\bigg(\frac{\d u}{\d t}\bigg)^2.
\end{split}                                        
\end{align}

Next we approximate the Lagrangian Eqn.~(\ref{eq:discretelagrangian}) as
\begin{align}
  \label{eq:continuouslagrangian}
\begin{split}
  L% &= \sum_{i=1}^{n} aT[\dot{\theta}(x_i,t)]- \sum_{i=1}^{n-1} a V[\theta(x_i,t)]\\
  % &\approx\int \d x \bigg\{T[\dot{\theta}(x,t)]-V[\theta(x,t)]\bigg\}\\
  % &=\int \d x \bigg\{ \frac{1}{2a}\frac{Mr^2}{r^2-u^2}\Big(\frac{\partial u}{\partial t}\Big)^2 \\
  % &~~~~~~~~~~~~~~~-\frac{2k}{a\overline{l}^2}(\frac{a^2}{2}\frac{\partial u}{\partial x}+\overline{u}^2-u^2)^2\bigg\}\\
  % &=\int \d x \bigg\{\frac{1}{2a}\frac{Mr^2}{r^2-u^2}\Big(\frac{\partial u}{\partial t}\Big)^2-\frac{ka^3}{2\overline{l}^2}\bigg(\frac{\partial u}{\partial x}\bigg)^2 \\
  % &~~~~~~~~~~~~~~~-\frac{2k}{a\overline{l}^2}(\overline{u}^2-u^2)^2-\frac{ka}{\overline{l}^2}\frac{\partial u}{\partial x}(\overline{u}^2-u^2)\\
  &\approx\int \d x \bigg\{\frac{M}{2a}\Big(\frac{\partial u}{\partial t}\Big)^2-\frac{ka^3}{2\overline{l}^2}\bigg(\frac{\partial u}{\partial x}\bigg)^2 \\
  &~~~~~~~~~~~~~~~-\frac{2k}{a\overline{l}^2}(\overline{u}^2-u^2)^2-\frac{ka}{\overline{l}^2}\frac{\partial u}{\partial x}(\overline{u}^2-u^2)\bigg\}.
\end{split}                                        
\end{align}
where we have taken the
leading order of the Taylor series expansion of the nonlinear kinetic term (in the
variable $u^2/r^2$), which is valid in the limit when $u \ll r$ or equivalently
$\sin \theta \ll 1$.

The first three terms in Eqn.~(\ref{eq:continuouslagrangian}) constitute the normal $\phi^4$ theory. The last term linear in
$\partial u/\partial x$, is an additional topological boundary term. Being a total derivative, it does not enter the Euler-Lagrange equation of motion and we
obtain the usual nonlinear Klein-Gordon equation 
\begin{align}
  \label{eq:phi4eqn}
\begin{split}
\frac{M}{a} \frac{\partial^2u}{\partial t^2}-\frac{ka^3}{\overline{l}^2}\frac{\partial^2u}{\partial x^2}-\frac{8k}{a\overline{l}^2}\overline{u}^2u+\frac{8k}{a\overline{l}^2}u^3=0,
\end{split}                                        
\end{align}
whose kink and antikink solutions are given by
\begin{align}
  \label{eq:kinksolution}
\begin{split}
u_0=\pm\overline{u}\tanh\Bigg[\frac{x-x_0-vt}{(a^2/2\overline{u})\sqrt{1-v^2/c^2}}\Bigg],
\end{split}                                        
\end{align}
where the $\pm$ denotes an (+)antikink and (-)kink respectively. Here, $v$ is the (anti)kink speed of propagation and $c=(a^2/\overline{l}\sqrt{k/M})$ is the speed of
sound in the medium. See Fig.~\ref{fig:akprofile} for a comparison with the discrete profile.

Note how the additional boundary term makes the potential energy density $V[\theta]$ a
perfect square, see Eqn.~(\ref{eq:continuouspotential}). For the kink configuration,
$V[\theta]$ therefore vanishes as is the case in the discrete topological chain. For the
antikink however, $V[\theta]$ is nonzero and is in fact twice of what we would expect in the normal $\phi^4$ theory (where both the kink and antikink configurations have the same energy). This is an agreement with our discussion on the discrete model in Section II.

\begin{figure}[h!]
  \centering
\captionsetup{justification=centerlast,margin=0cm,singlelinecheck=false}
  \includegraphics[width=0.45\textwidth]{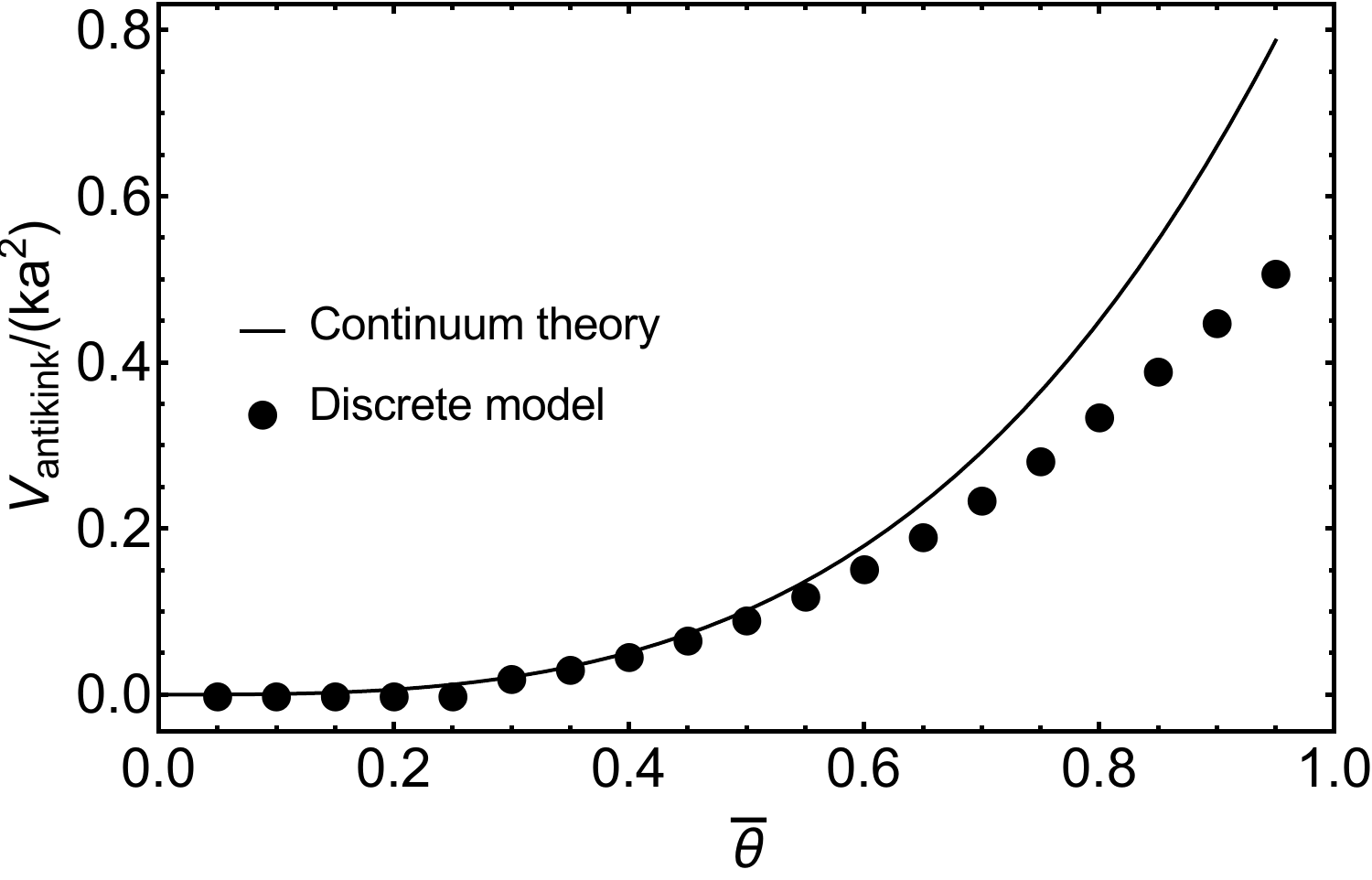}
  \caption{The normalized potential energy plotted against the equilibrium angle $\overline{\theta}$, for a static antikink configuration in a topological rotor chain with with $r/a=0.8$. The discrete model has 60 rotors. Note that the wobbler transition~\cite{Chen2014} is around $\overline{\theta}=\sin ^{-1}\left(\frac{a}{2 r}\right)=0.67$, which is close to where
the continuum theory starts to significantly deviate from the discrete model.}
  \label{fig:akenergy}
\end{figure}

Upon substituting the static~($v=0$) antikink profile from Eqn.~(\ref{eq:kinksolution}) into Eqn.~(\ref{eq:continuouslagrangian}) and completing the integral, we obtain the potential energy of the topological rotor chain with an antikink profile
\begin{align}
  \label{eq:akenergy}
  \begin{split}
    V_{antikink}/(ka^2)=\frac{16}{3}\frac{(r/a)^3\sin^3\overline{\theta}}{1+4(r/a)^2\cos^2\overline{\theta}}.
  \end{split}
\end{align}
In Fig.~\ref{fig:akenergy}, we compare this expression with the predictions from the discrete model. We see that the continuum theory agrees reasonably well with the discrete
model as long as $\overline{\theta}$ is less than approximately 0.6, below which, the width of the antikink is larger than the lattice spacing and therefore, a continuum approximation well justified.

\section{Linear mode analysis: tangent stiffness matrix approach}

 We now study small oscillations around the kink and antikink configurations, first in the
 continuum limit, and next in the discrete model by developing the tangent stiffness matrix approach.  In the continuum limit, we make the ansatz $u=u_0+\delta u$ and substitute into Eqn.~(\ref{eq:phi4eqn}) retaining only terms linear in $\delta u$:
\begin{align}
  \label{eq:smallosci}
  \begin{split}
    \frac{M}{a}\frac{\partial^2\delta u}{\partial
      t^2}-\frac{ka^3}{\overline{l}^2}\frac{\partial^2\delta u}{\partial
      x^2}-\frac{8k}{a\overline{l}^2}(\overline{u}^2-3u_0^2)\delta u=0
\end{split}                                        
\end{align}
If we Fourier transform Eqn.~(\ref{eq:smallosci}) with respect to time, we obtain a Sch\"{o}dinger-like equation with a solvable potential~\cite{Campbell1983,Makhankov1990}. This yields one continuous spectral band as well as two discrete modes -- one translation mode for the (anti)kink and one shape mode, which corresponds to small deformations of the shape of the (anti)kink localized around the center of their profile. For the topological rotor chain, the frequencies of the two discrete modes are:

\begin{align}
  \label{eq:continuumtheoryfrequency1}
  \begin{split}
    \omega_t&=0, ~~\textrm{for the translation mode}\\
  \end{split}                                        
\end{align}
\begin{align}
  \label{eq:continuumtheoryfrequency2}
  \begin{split}
    \omega_s&=(r/a)\sqrt{12k/M}\sin\overline{\theta}/\sqrt{1+4(r/a)^2\cos^2\overline{\theta}},\\
  &~~~~~~~~~~~~~~~~~~~\textrm{for the shape mode}.
  \end{split}                                        
\end{align}

\begin{figure}[h!]
  \subfloat[]{\label{fig:kinkconft}
    \includegraphics[width=0.45\textwidth]{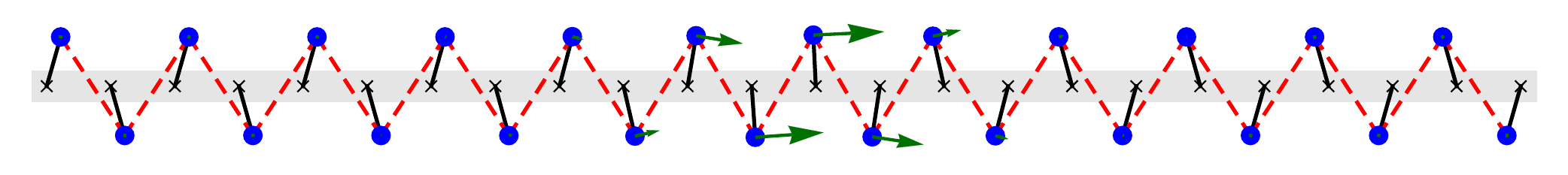}}
  
  \subfloat[]{\label{fig:kinkconfs}
    \includegraphics[width=0.45\textwidth]{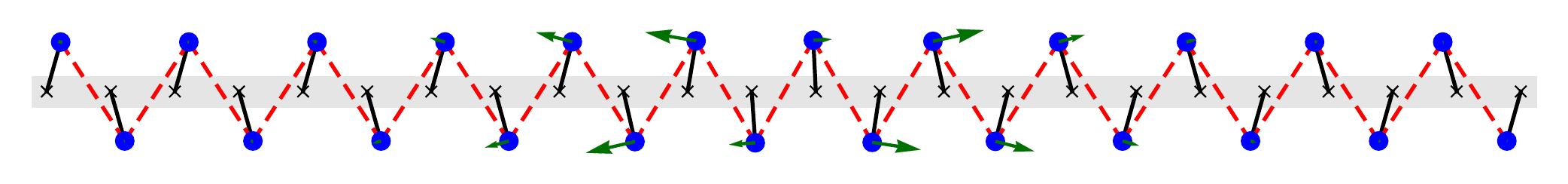}}
  
  \subfloat[]{\label{fig:antikinkconft}
    \includegraphics[width=0.45\textwidth]{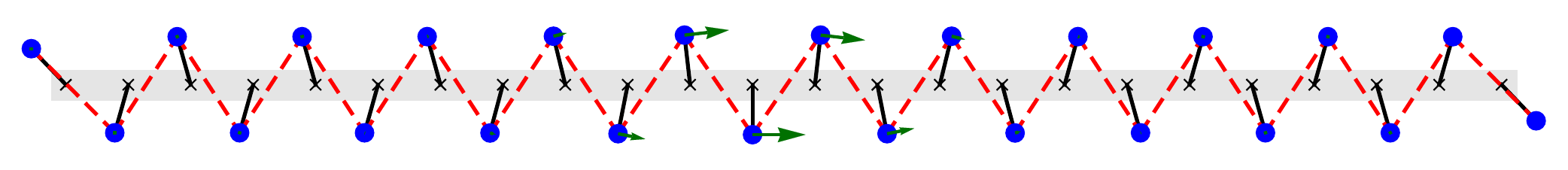}}
  
  \subfloat[]{\label{fig:antikinkconfs}
  \includegraphics[width=0.45\textwidth]{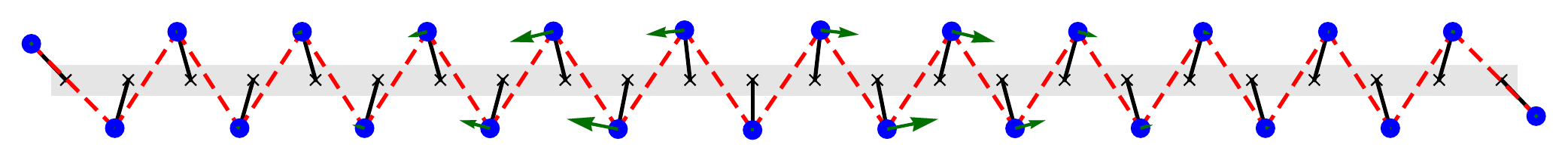}}
  \caption{\label{fig:mode-config}The configurations of (\textbf{a})~the kink translation mode, (\textbf{b})~the
    kink shape mode, (\textbf{c})~the antikink translation mode and (\textbf{d})~the
    antikink shape mode. The green arrows depict the mode component of each
    rotor.}
\end{figure}

In Fig.~\ref{fig:kinkconft} and \ref{fig:antikinkconft}, the kink and antikink are located in the middle of the chain. The mode arrows (in green) that all point in the same direction, correspond to a translation mode. In Fig.~\ref{fig:kinkconfs} and \ref{fig:antikinkconfs}, the arrows on either side of the (anti)kink, point in opposite directions and these correspond to shape deformations of the (anti)kink.

\begin{figure}[h!]
  \subfloat[]{\label{fig:w-th-k}
    \includegraphics[width=0.45\textwidth]{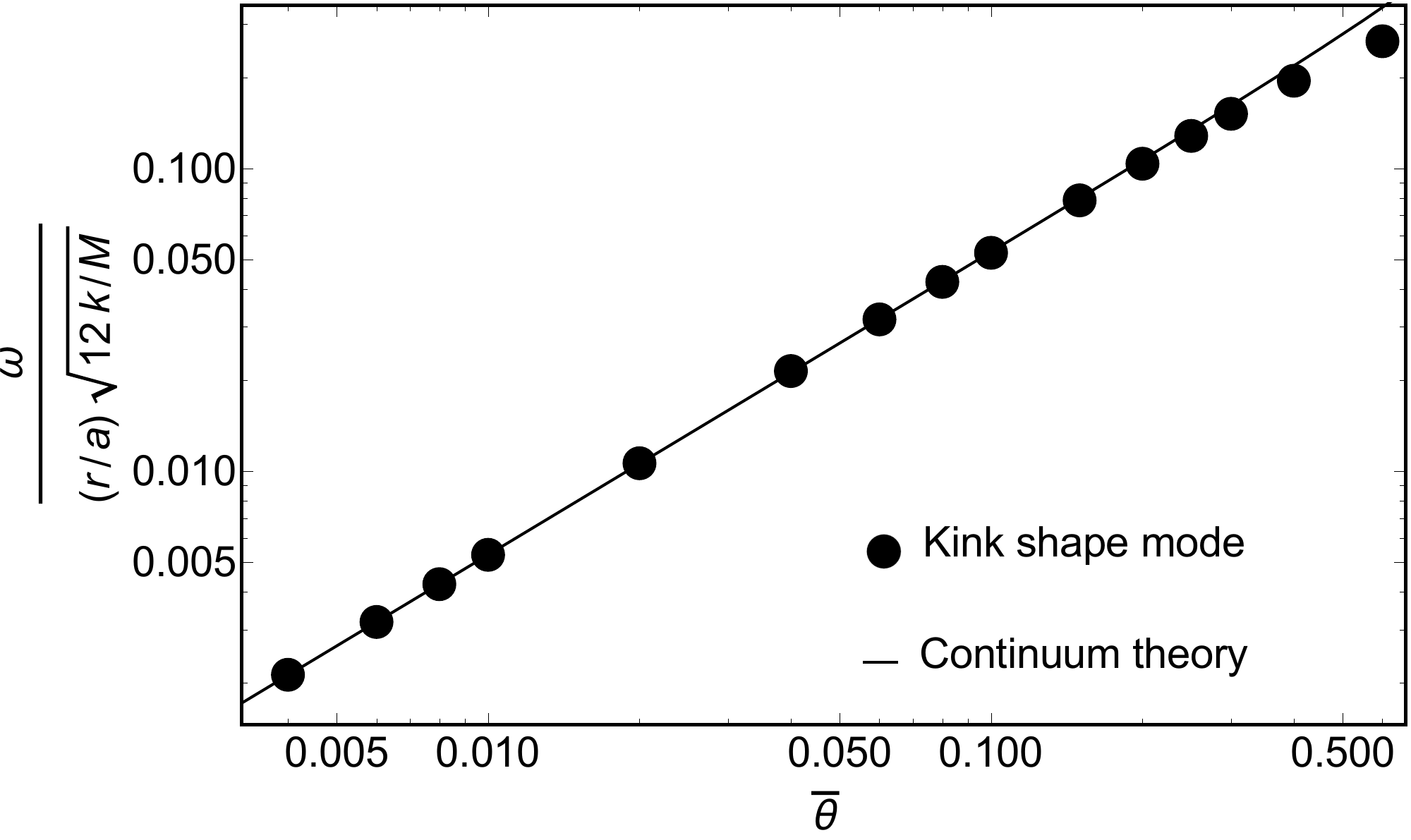}}
  
  \subfloat[]{\label{fig:w-th-ak}
  \includegraphics[width=0.45\textwidth]{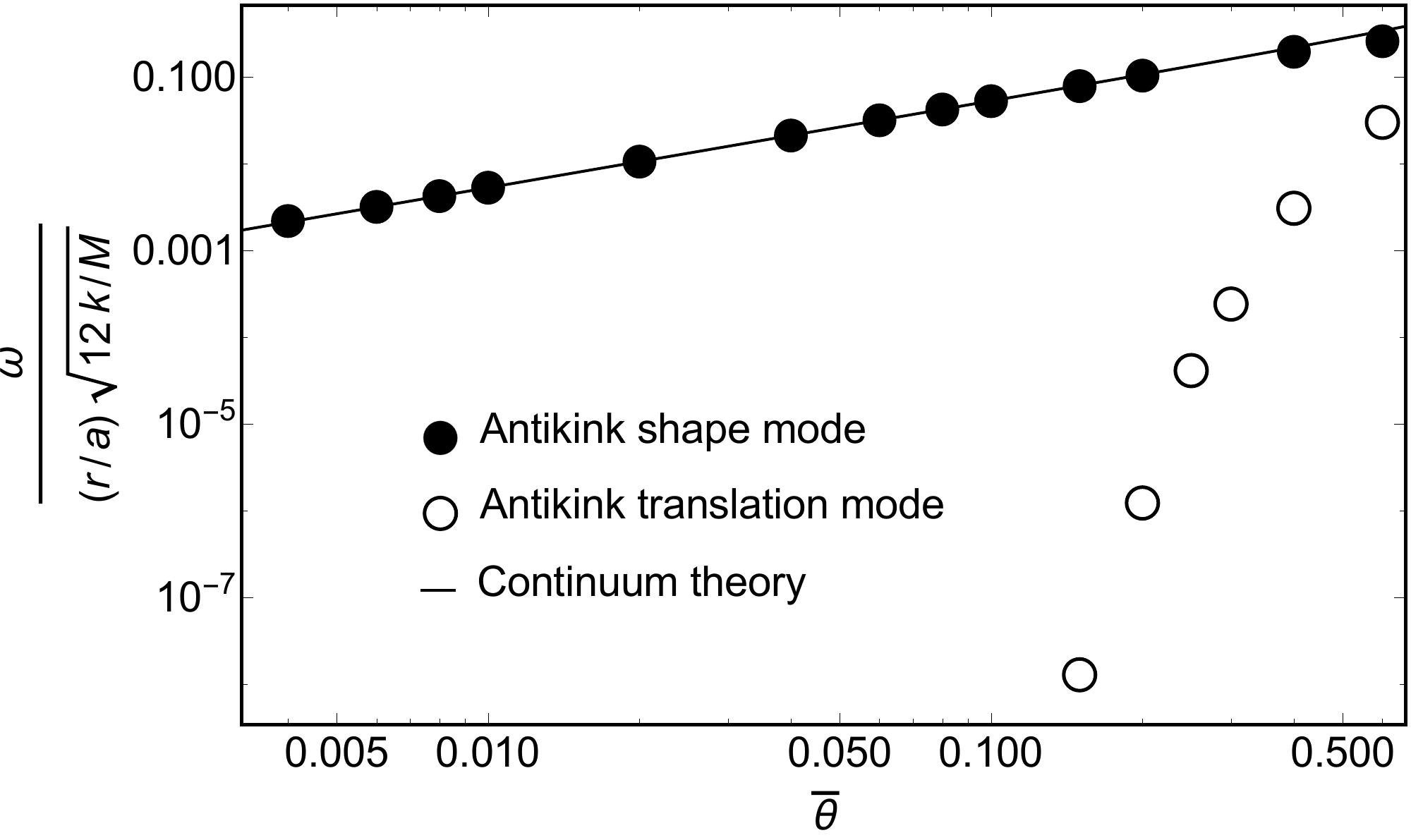}}
  \caption{\label{fig:w-th}The frequencies $\omega$ of localized mode(s) for (\textbf{a})~the kink and (\textbf{b})~the antikink as a function of $\overline{\theta}$ for a rotor chain with $r/a=0.8$. The data points are numerically obtained from the tangent stiffness matrix approach, filled circles correspond to the shape mode ($\omega _s$), while open circles correspond to the translation mode ($\omega _t$) . The curves are from the continuum theory. The frequencies for the kink translation mode for all $\bar{\theta}$ and the frequencies for the antikink translation mode for $\overline{\theta}<0.1$ are effectively zero at machine precision and thus, not visible in the figure. }
\end{figure}

In Appendix~\ref{App: AppendixD}, we follow the approach proposed by
Guest~\cite{Guest2006} to derive the \textit{tangent stiffness matrix}~$\mathbf{K}$ for
prestressed mechanical structures. With $\mathbf{K}$ we numerically obtain the frequencies
of localized modes for the discrete chain model and compare them with
the predictions of
the continuum theory (Eqn.~(\ref{eq:continuumtheoryfrequency1}) and
Eqn.~(\ref{eq:continuumtheoryfrequency2})) in Fig.~\ref{fig:w-th}. We find that the
translation mode $\omega _t$ for the kink indeed vanishes (within machine-precision in our
numerics) for all values of $\overline{\theta}$ and is thus absent in the range of the
log-log plot shown in Fig.~\ref{fig:w-th-k}). However, as seen in Fig.~\ref{fig:w-th-ak},
the translation mode (open circles) for the antikink is nonzero.

For the shape mode $\omega _s$ (filled circles), we find the numerical
results for both the kink and antikink to be in good agreement
with the continuum theory at small $\overline{\theta}$. Note that in Fig.~\ref{fig:w-th-ak}, although the antikink has a finite nonzero $\omega_t$, the value is still significantly smaller than $\omega_s$.

\section{Kink/antikink propagation in ordered lattices}
In the previous section, we have seen that for the discrete
topological chain, the energy of the translation mode for the kink is zero, whereas
that for the antikink is non-zero. Note that the standard discretization of a $\phi^4$ field theory leads to a non-zero translation mode for both the kink and antikink ~\cite{Roy2007}. Thus, the kink here differs qualitatively from the
antikink in that it has a zero mode even when we consider the discrete model. We next numerically simulate the propagation of a kink and antikink along the discrete chain and see how this difference manifests in their dynamics. 

We numerically integrate Newtons equation of motion for the rotors using molecular
dynamics simulations. (The simulation settings are described in Appendix \ref{App: AppendixB}.) A
stable chain configuration with a single kink or antikink is used as the initial
configuration (see Figs.~\ref{fig:kinkconft}-~\ref{fig:antikinkconft} for the initial
conditions used). An excitation is set in motion with a velocity along the direction of
the translation mode, but with variable amplitudes.

\begin{figure}[h!]
  \centering
\captionsetup{justification=centerlast,margin=0cm,singlelinecheck=false}
  \includegraphics[width=0.45\textwidth]{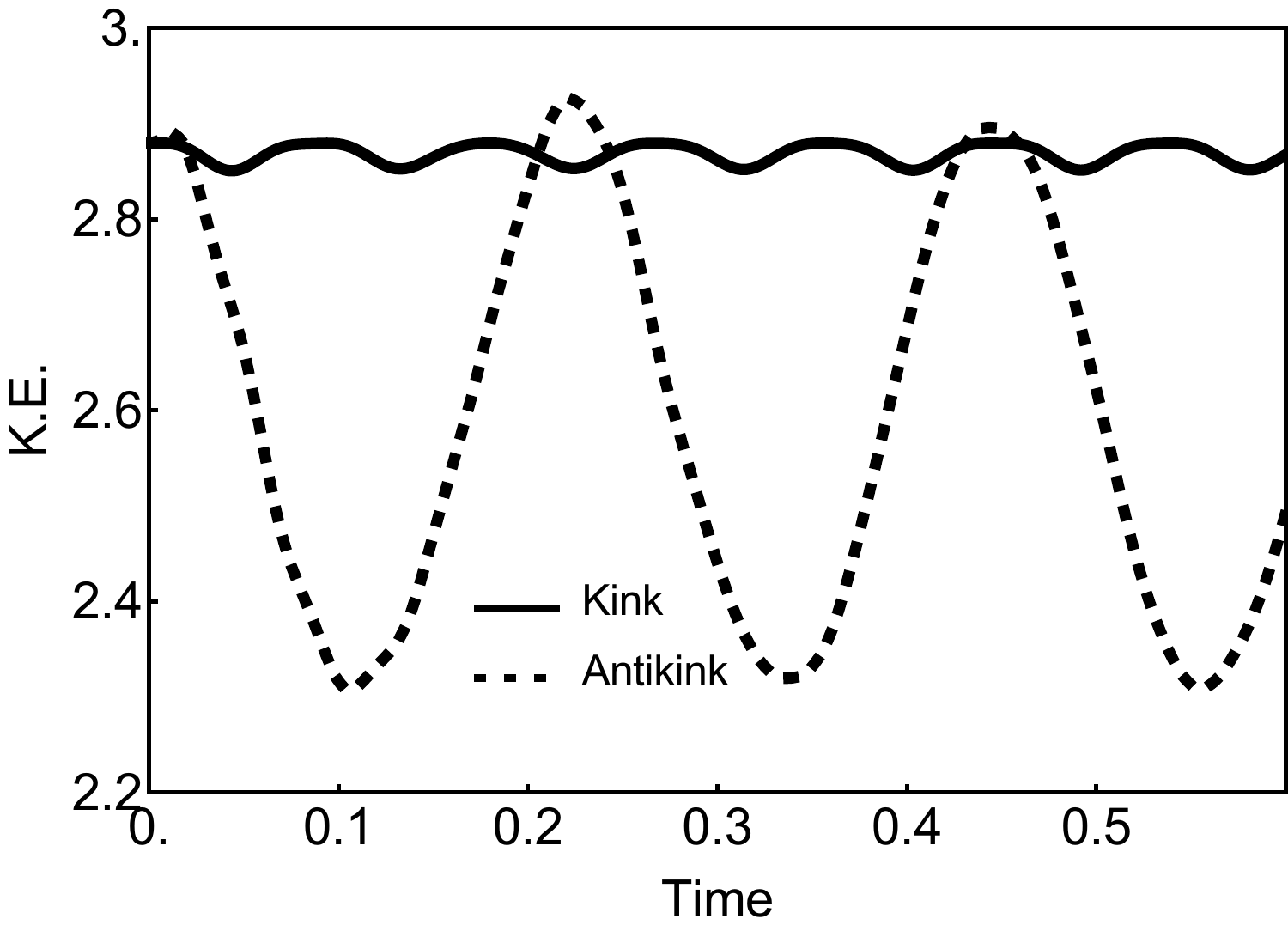}
  \caption{Time evolution of the kinetic energy for a kink (Fig.~\ref{fig:kinkconft}) and an antikink (Fig.~\ref{fig:antikinkconft}) in a topological rotor chain with non-dimensional parameters $M=1$, $k=10000$, $r/a=0.8$, $\overline{\theta}=0.58$. The magnitude of initial velocity in both cases is $v_0=2.4$. The units of energy and velocity are determined by the aforementioned physical parameters. The kink propagation only results in small oscillation of the K.E.~whereas we see significant fluctuations during the propagation of an antikink. These can be traced to the Peierls-Nabarro potential as shown in Fig.~\ref{fig:akconfigpn}}
  \label{fig:ketimeplot}
\end{figure}

In Fig.~\ref{fig:ketimeplot}, we plot the kinetic energy~(K.E.) of the chain as a function of time for a set of parameters, for a kink excitation (solid curve) and an antikink excitation (dashed curve). The K.E.~of the kink remains nearly constant for all times with some small fluctuations (as the springs have to slightly deform to transport energy by simultaneously minimize the potential and kinetic energy). However in comparison, the K.E.~of the antikink for the same set of initial parameters changes significantly as it propagates down the chain. The key point is that the kink and antikink do not propagate in the same way. 

The asymmetry between a {\em static} kink and antikink configuration was discussed in~\cite{Chen2014}. Further, we also know from Eqn.~(\ref{eq:continuouslagrangian}) (and the ensuing discussion) that in the continuum limit,  the topological rotor chain is approximately described by a $\phi^4$ theory with an additional topological boundary term which ensures that the potential energy of the kink is zero while that for the antikink is nonzero (see Ref. \cite{Vitelli2014} for an interpretation of this fact in terms of supersymmetry breaking).
However, the additional boundary term does not affect the continuum equation of motion and thus, both the kink and antikink should have translational invariance in this limit and their dynamics should not have differed.

\begin{figure}[h!]
  \subfloat[]{\label{fig:akconfigpn1}
    \includegraphics[width=0.45\textwidth]{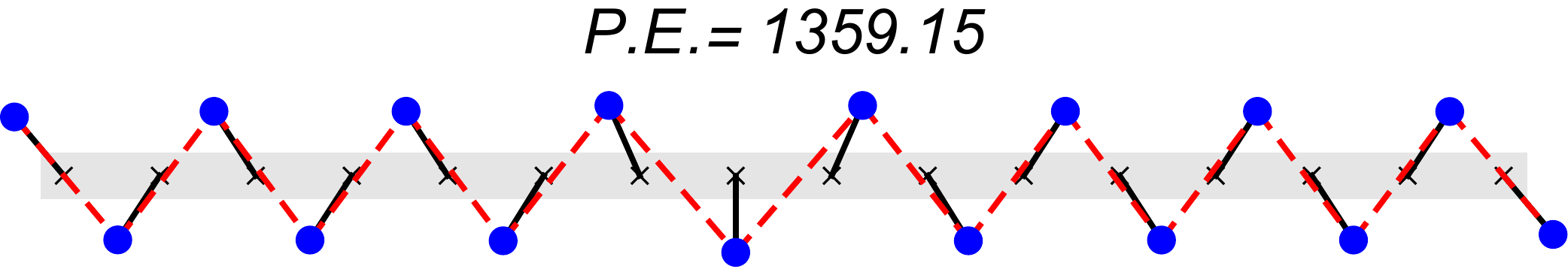}}
  
  \subfloat[]{\label{fig:akconfigpn2}
  \includegraphics[width=0.45\textwidth]{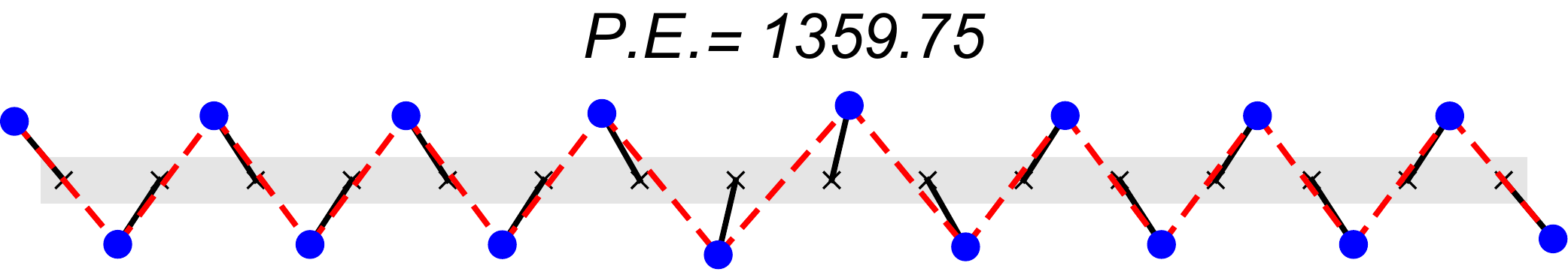}}
  \caption{\label{fig:akconfigpn}Two equilibrium configurations in the potential energy landscape of a static antikink: (\textbf{a})~a minimum and (\textbf{b})~a saddle point, respectively. The topological chain has the same configuration parameters as in Fig.~\ref{fig:ketimeplot}. }
\end{figure}

The reason for this asymmetrical behavior can be understood only if we examine the discrete model. The system with free boundary conditions has $n$ rotors and $n-1$ springs, and the static kink does not require any of the springs to be stretched. We can therefore interpret the springs as constraints. Thus, the discrete kink's equilibrium manifold is a continuous curve embedded in the $n$-dimensional configuration space of the rotor angles $\theta_i$ and the kink can be positioned stably anywhere along the chain. By contrast, an antikink requires the springs to be stretched. Forces on each of the rotors have to be balanced for the system to be in mechanical equilibrium. So the possible equilibrium configurations have to be symmetrical locally around the center of the antikink, as shown in Fig.~\ref{fig:akconfigpn}. As a result, the equilibrium manifold for an antikink is not a continuous curve but rather, consists of a set of discrete points. These correspond to either saddle points or minima in the potential landscape. Any locally asymmetrical configuration is therefore not stable and will slide towards a minima.

The saddle points and their nearest minima can be connected by an ``adiabatic trajectory''~\cite{Braun1998}, which is a curve of steepest descent. The concept of an
adiabatic trajectory is useful in two ways. First, it describes the slow motion of the antikink through the chain. The position of the antikink center can be defined by a coordinate along such a trajectory. Secondly, it helps to rigorously define the so-called Peierls-Nabarro (PN) potential~\cite{Combs1983,Braun1998,Roy2007}, which is the effective periodic potential that the antikink feels as it moves along the adiabatic trajectory. A saddle point in the full potential energy landscape corresponds to a maximum along the adiabatic trajectory (while a minimum is still a minimum). Note that although the antikink's K.E.~fluctuations in Fig.~\ref{fig:ketimeplot} do not strictly equal its PN potential barrier, the former reveals the existence of the latter.

In Appendix~\ref{app:AppendixPN}, we derive the PN potential barrier from the continuum
theory
\begin{align}
  \label{eq:PNB2}
  \begin{split}
    V_{PNB}% &=\int_{-\infty}^{+\infty}\d n'\frac{16k\overline{u}^4}{\overline{l}^2}\sech^4\Big(\frac{n'a}{w}\Big)
    % e^{-2\pi in'}\\
    &=\frac{4\pi^2\big(\pi^2+(a/w)^2\big)}{3\big(1+4(r/a)^2-(a/w)^2\big)\sinh(\pi^2w/a)}\\
    &\propto e^{-\pi^2w/a}~~~~~~~~~~~~\textrm{for large $w/a$.}
  \end{split}
\end{align}
This shows that the PN barrier decays exponentially as the width $w$ of the antikink
increases.

\begin{figure}[h!]
  \centering
\captionsetup{justification=centerlast,margin=0cm,singlelinecheck=false}
    \includegraphics[width=0.45\textwidth]{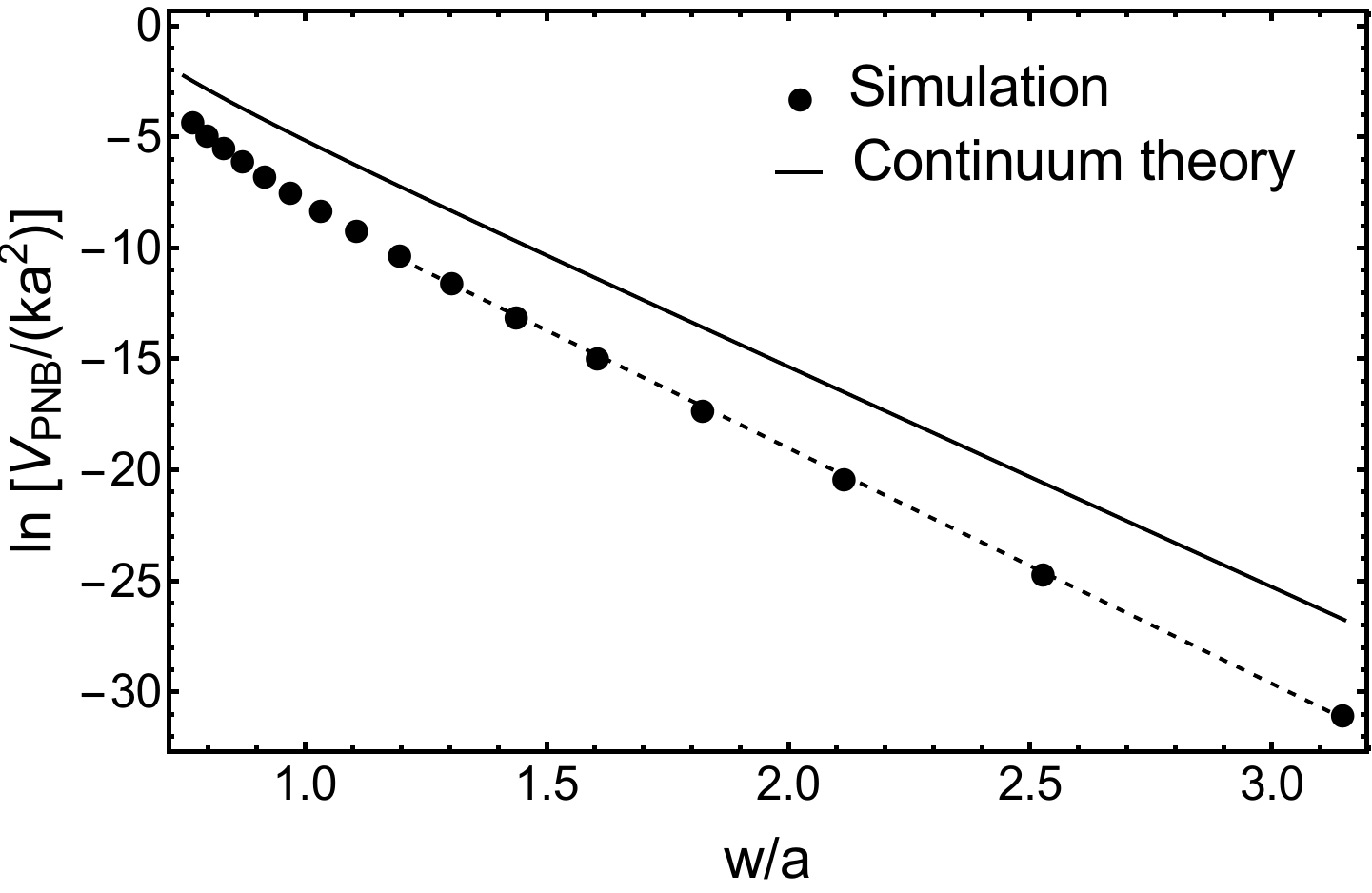}
    \caption{The dependence of the normalized PN barrier~($V_{PNB}/ka^2$) on the normalized antikink width~($w/a$), for both the discrete model (black circles) and the continuum theory (solid line). The slope of the dashed line (fit to simulation) is $-10.6$, in reasonable agreement with the predictions from the continuum theory in Eqn.~(\ref{eq:PNB2}), which gives a slope $-\pi^2 \approx -9.9$. }
    \label{fig:logvpn-theta}
\end{figure}

We next compare the theoretical results with numerical simulations. We obtain the exact PN
barrier by computing the difference in potential energy between the two types of
equilibrium points: a minima and a saddle point, see Fig.~\ref{fig:akconfigpn}, where for
a given set of parameters,  we find the barrier height to be $1359.75-1359.15=0.60$,
consistent with the magnitude of the K.E.~fluctuations shown in Fig.~\ref{fig:ketimeplot}
for the same set of parameters. By repeating this calculation for systems with various
antikink widths $w$, we obtain the dependence of the normalized PN barrier
$V_{PNB}/(ka^2)$ on $w/a$, which we show in Fig.~\ref{fig:logvpn-theta}. We compare these
with the predictions from the continuum theory, given by Eqn.~(\ref{eq:PNB2}). The
numerical results (filled circles) obtained from the discrete lattice and the theoretical
predictions (continuous curve) follow a similar trend, but differ by
at least one order of magnitude. This can be explained by the fact
that the discreteness of the lattice is ignored in the theory when we take the continuum limit in going from Eqn.~(\ref{eq:discretepotential2}) to Eqn.~(\ref{eq:continuouspotential}). See~\cite{Combs1983} for a thorough discussion of the effect of lattice discreteness on the single-kink dynamics in a $\phi^4$ model.

\begin{figure}[h!]
  \centering
\captionsetup{justification=centerlast,margin=0cm,singlelinecheck=false}
    \includegraphics[width=0.45\textwidth]{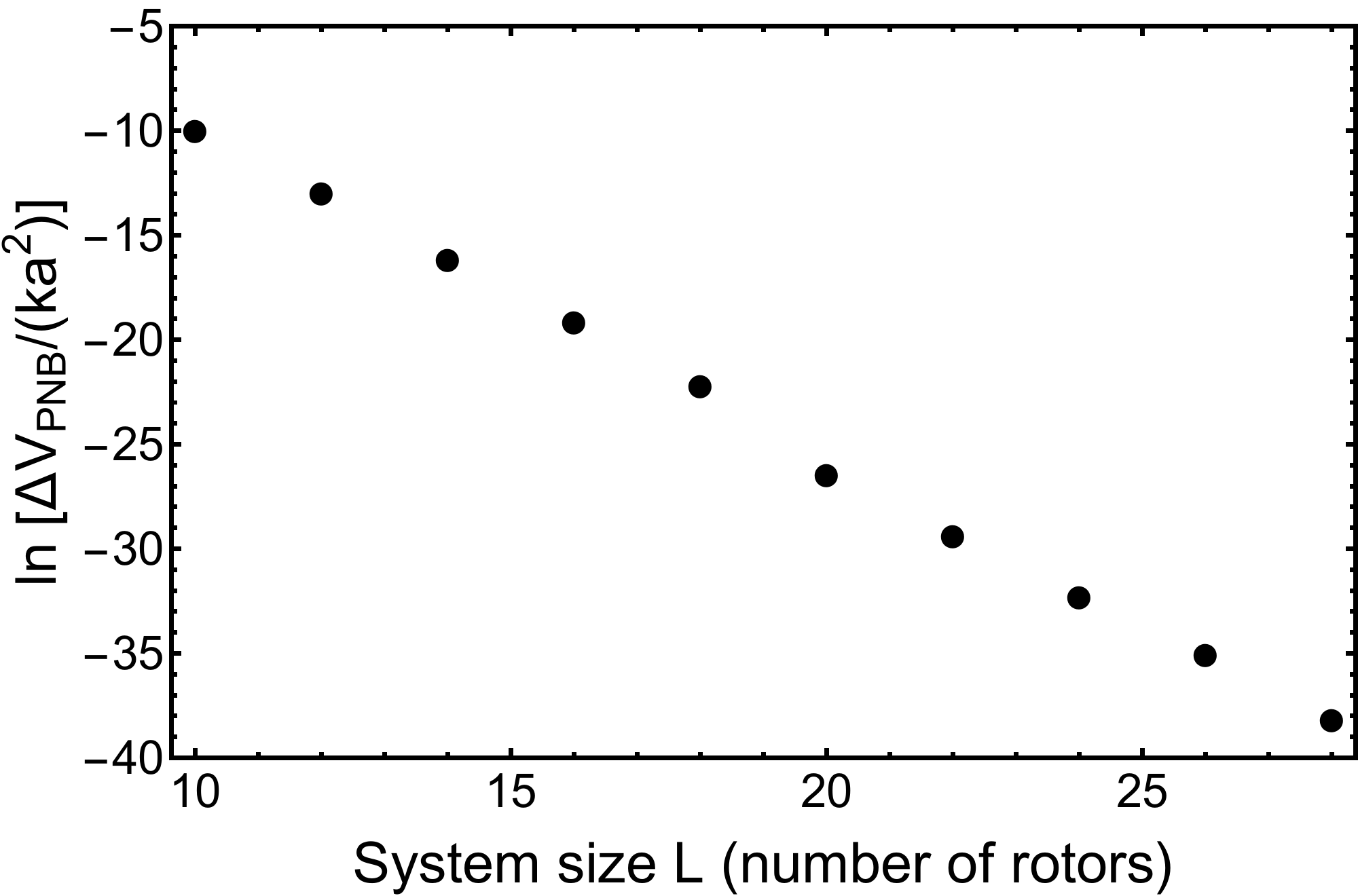}
    \caption{The finite-size effect on $V_{PNB}$. $\Delta V_{PNB}$ is defined as $V_{PNB}(L) - V_{PNB}(L=60)$.
      The configuration parameters are $r/a=0.8$ and $\overline{\theta}=0.40$. }
    \label{fig:logvpn-size}
\end{figure}

Further, we also investigate finite-size corrections to the PN barrier, or more precisely, the difference between $V_{PNB}$ for a system with a small finite size and that for a system with a sufficiently larger size~(60 rotors). We find that finite size effects decay quickly as an exponential function with increasing system size for a topological rotor chain with a central antikink (see Fig.~\ref{fig:logvpn-size}). This is because an antikink configuration is a localized object. The components of its displacement, its translation mode, as well as its shape mode, decay exponentially away from its center and therefore, so does the effect of any boundaries.

\begin{figure}[h!]
  \centering
\captionsetup{justification=centerlast,margin=0cm,singlelinecheck=false}
  \includegraphics[width=0.4\textwidth]{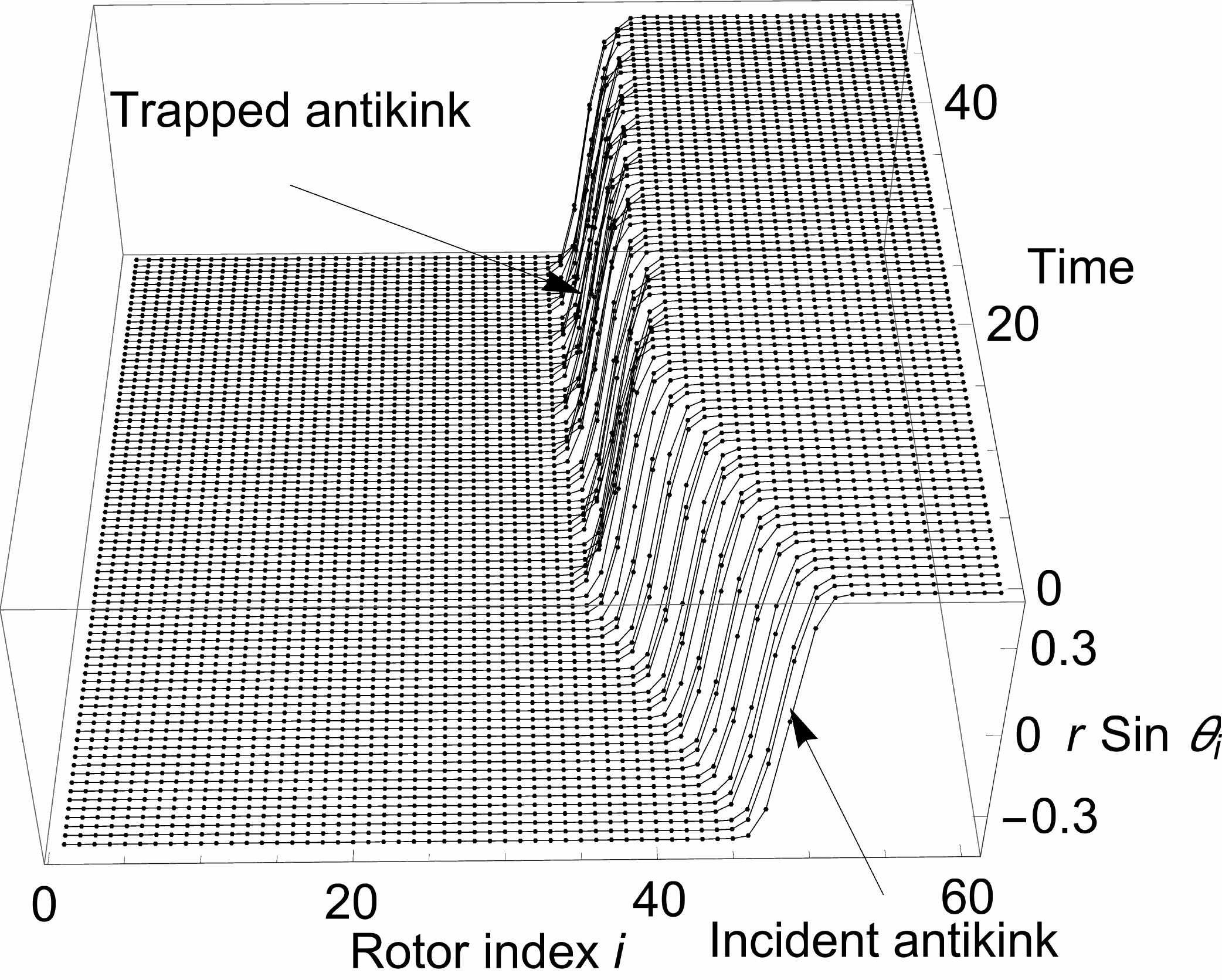}
  \caption{Perspective view of a moving antikink trapped in its Peierls-Nabarro barrier around $ \textrm{Time} =20$ near Rotor \#35. The topological rotor chain has the same configuration parameters as in Fig.~\ref{fig:ketimeplot} and the initial antikink velocity is $v_0=1.1$ in non-dimensional units.}
  \label{fig:akpntrapprofile}
\end{figure}

To summarize, for the topological rotor chain that we study, the PN barrier for a kink vanishes and that for an antikink is finite. This, not only affects how their respective kinetic energies fluctuate over a lattice spacing, but also affects their dynamics over long distances. It is well known that $\phi^4$  kinks and antikinks are non-integrable solutions ~\cite{Makhankov1990}. Although the kinks and antikinks are ``topologically'' robust objects, they still tend to dissipate energy into phonons and into shape fluctuations as they propagate. Once an antikink has lost too much kinetic energy to be able to overcome the PN barrier, it gets trapped in a PN potential minimum, as shown in Fig.~\ref{fig:akpntrapprofile}. On the other hand, for the topological rotor chain that we study, the kink never gets trapped, since its PN barrier vanishes.

\section{Effect of spring stiffness impurities}

We next numerically explore whether the kink-antikink asymmetry also manifests in the way
these excitations interact with a single lattice impurity, a natural starting point to
study their propagation in disordered lattices. For the conventional $\phi^4$ models,
previous studies on kink-impurity interactions (in both discrete models~\cite{Fraggis1989}
and continuum field models~\cite{Fei1992}) have shown that scattering can result in
transmission, trapping or reflection of kinks, depending on the type of the impurity, the
attraction/repulsion strength of the impurity and the kink's initial velocity. Although
similar scattering also occurs in the topological rotor chain model, we also find other
novel phenomena, for instance, the kink can split into two kinks and one antikink.
Moreover as we will see, kinks and antikinks no longer scatter in the same way -- a
feature which underscores the kink-antikink asymmetry in our topological rotor chain. In
this work, we study impurities in properties of the springs, which yield a richer set of
effects on the response than mass impurities.

In this section, we model an impurity by changing the spring stiffness constant at a
single site (Fig.~\ref{fig:configuration}). We study a topological chain with lattice
spacing $a=1$ and rotor length $r/a=0.8$ and with equillibirum angle
$\overline{\theta}=0.28$. We perform Newtonian dynamics simulation on a system with 60
rotors using free boundary conditions, and for a range of impurity spring stiffness
constant $k_i$ and kink/antikink initial velocity $v_0$. See
Fig.~\ref{fig:scheme-scattering} for a table of the possible scattering scenarios that we
observe.

\begin{figure}[h!]
  \subfloat[]{\label{fig:k-transmit-utn}
    \includegraphics[width=0.4\textwidth]{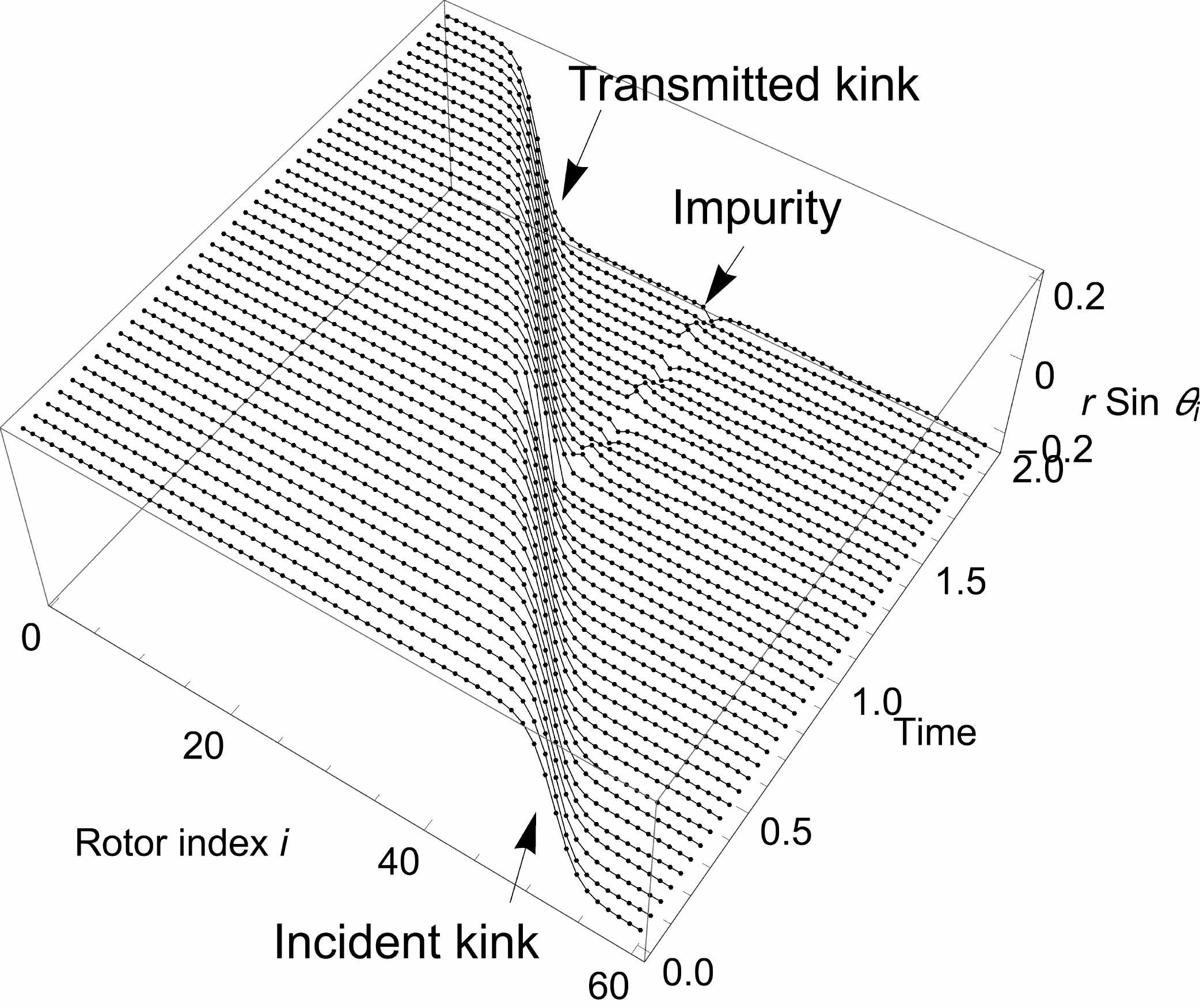}}
  
  \subfloat[]{\label{fig:k-split-utn}
  \includegraphics[width=0.4\textwidth]{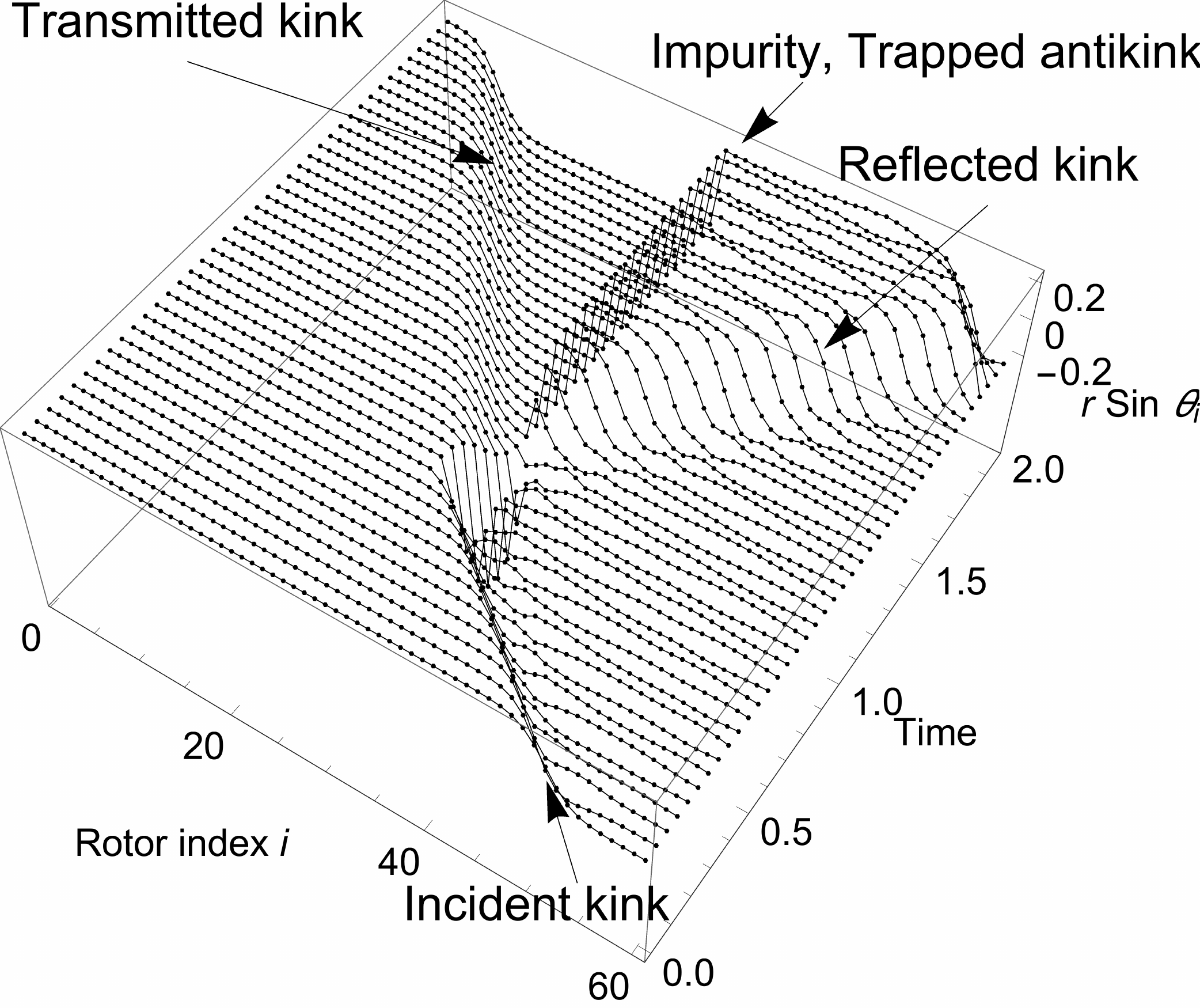}}
  \caption{\label{fig:k-scatter}A kink interacts with an impurity (different spring stiffness) and is either (\textbf{a})~ transmitted, shown here for $v_0=4.0$ and $k_i/k=0.10$ or (\textbf{b})~splits into a transmitted kink, a reflected kink and an antikink trapped at the impurity, shown here for $v_0=9.6$ and $k_i/k=0.01$. The non-dimensional parameters are $M = 1, k = 10000, r/a = 0.8, \bar{\theta} = 0.28$.}
\end{figure}

Consider first the kink-impurity interaction. For most $k_i$ and $v_0$, the kink simply passes through the impurity and may excite an impurity mode, which can be seen in the form of small fluctuations in the middle of the chain as shown in Fig.~\ref{fig:k-transmit-utn}. When the impurity spring is sufficiently soft, the incident kink splits into three: a transmitted kink, an antikink that is trapped at the impurity and a reflected kink. This is shown in Fig.~\ref{fig:k-split-utn}.

\begin{figure}[h!]
  \subfloat[]{\label{fig:ak-transmit-utn}
    \includegraphics[width=0.4\textwidth]{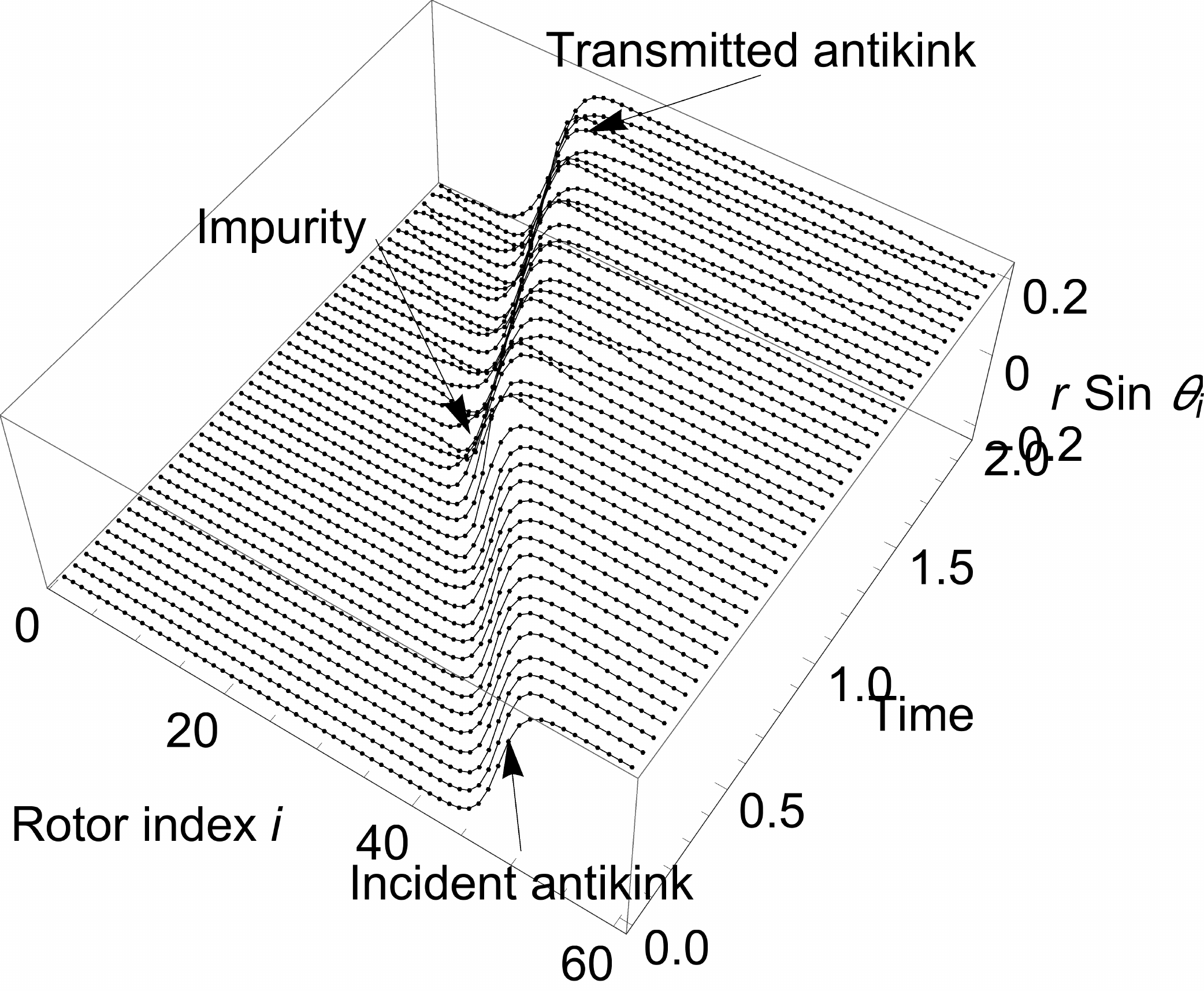}}
  
  \subfloat[]{\label{fig:ak-trap-utn}
    \includegraphics[width=0.4\textwidth]{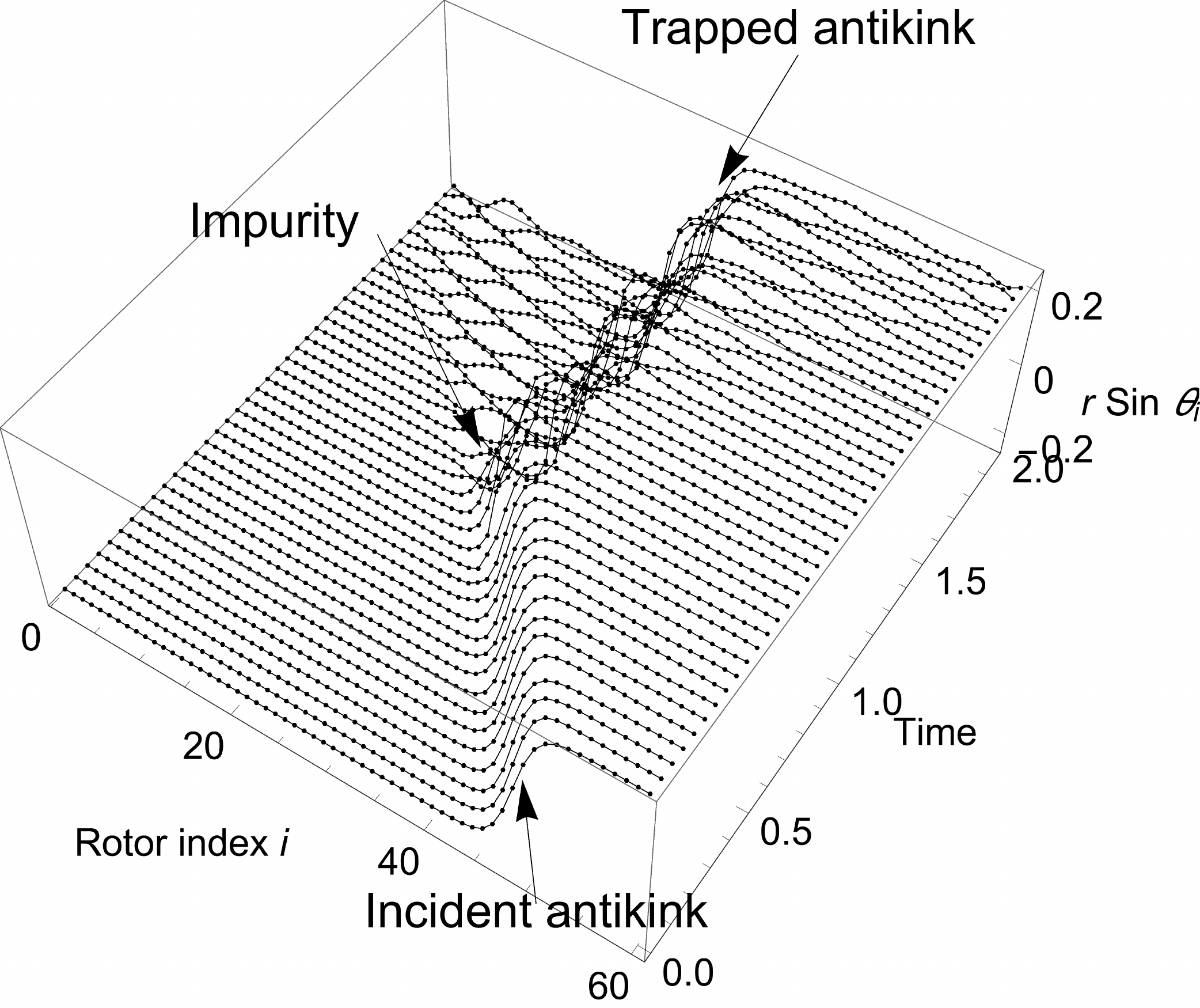}}
  
  \subfloat[]{\label{fig:ak-reflect-utn}
    \includegraphics[width=0.4\textwidth]{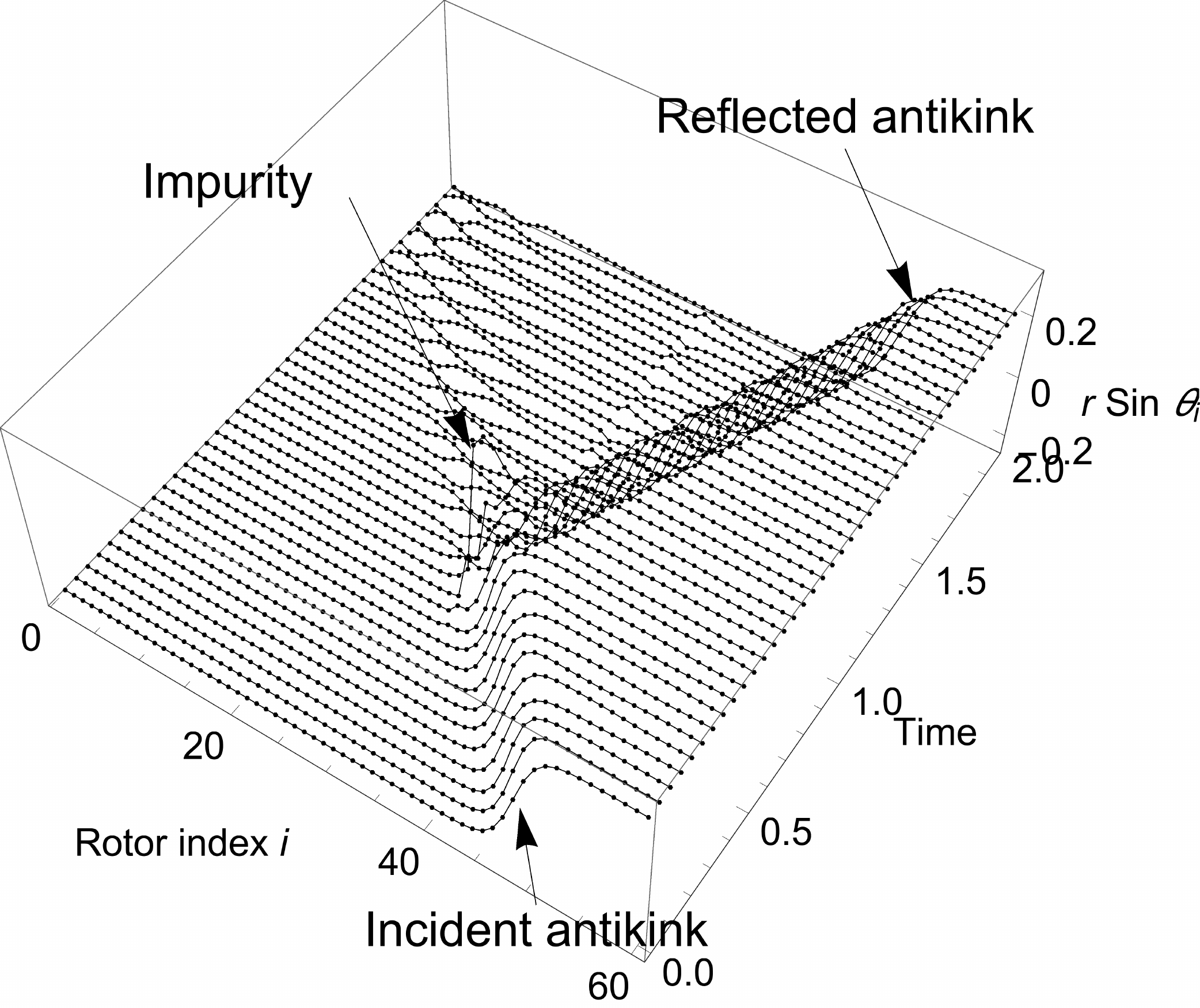}}
  \caption{\label{fig:ak-scatter}An antikink interacts with an impurity and is either (\textbf{a})~transmitted, shown here for $v_0=4.0$ and $k_i/k=0.80$, (\textbf{b})~trapped, shown here for $v_0=4.0$ and $k_i/k=0.70$ or (\textbf{c})~reflected, shown here for  $v_0=4.8$ and $k_i/k=0.20$. The system parameters are the same as Fig.~\ref{fig:k-scatter}.}
\end{figure}

Antikink scattering results in an ever richer set of behaviors. Recall that the springs
near the location of an antikink are always stretched significantly, see
Fig.~\ref{fig:akspringstretch}. For $k_i/k$ near 1, the antikink gets transmitted with
energy dissipation and thus slows down~(Fig.~\ref{fig:ak-transmit-utn}). Softening the
impurity spring stiffness creates an attractive potential well for the antikink. The
antikink may then release a part of its potential energy and get trapped at such an
impurity site (Fig.~\ref{fig:ak-trap-utn}). If the impurity spring is made even softer,
such that an antikink can no longer transfer its kinetic energy forward or dissipate it
sufficiently quickly to be trapped, the incident antikink is completely reflected
(Fig.~\ref{fig:ak-reflect-utn}). For similar reasons, a stiffer impurity acts like a
repulsive potential well that can reflect slow moving antikinks.

\begin{figure}[h!]
  \subfloat[]{\label{fig:phasediagram-kink}
    \includegraphics[width=0.5\textwidth]{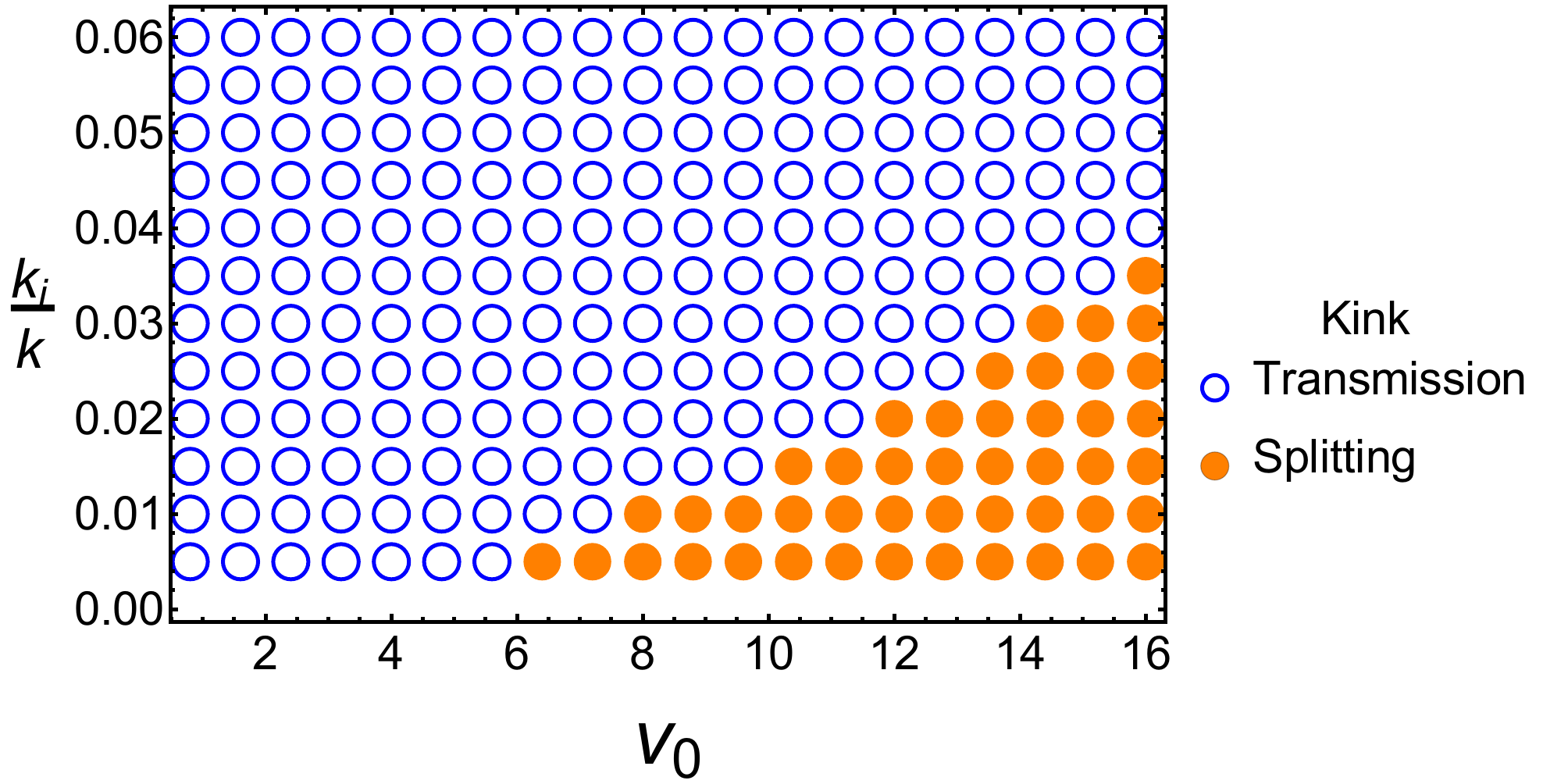}}
  
  \subfloat[]{\label{fig:phasediagram-antikink}
  \includegraphics[width=0.5\textwidth]{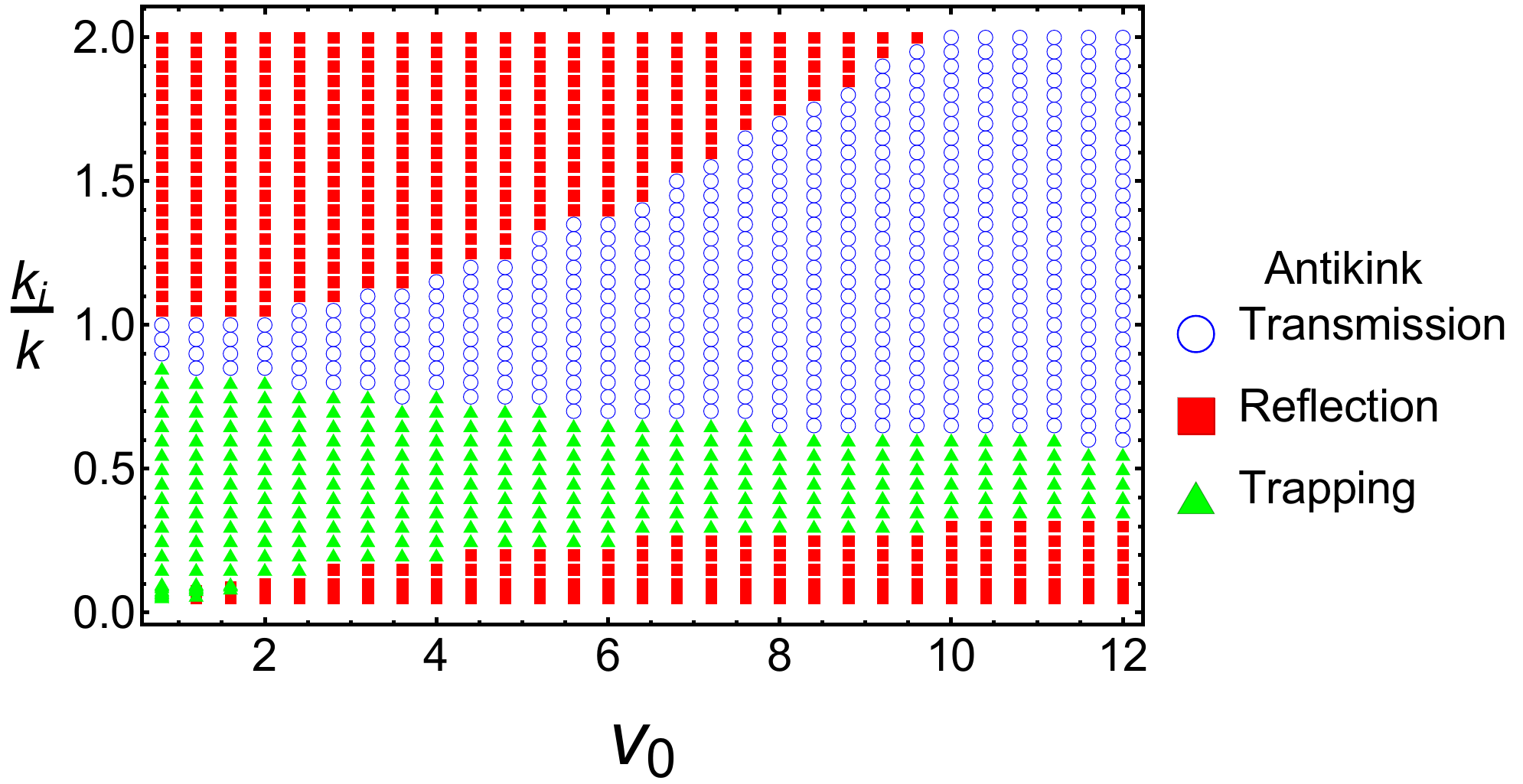}}  
\caption{\label{fig:phasediagram}The phase diagram of the scattering behavior in the
  parameter space of normalized spring constant of impurity $k_i/k$ and kink initial
  velocity $v_0$ for (\textbf{a})~the kink and (\textbf{b})~the antikink. The system
  parameters are the same as Fig.~\ref{fig:k-scatter}. The lower limit of $v_0$ for the
  antikink is around 0.7, below which even the PN barrier in a perfect chain will capture
  the antikink. }
\end{figure}

These numerical results are summarized in the phase diagrams in the space of $k_i$ and
$v_0$ in Fig.~\ref{fig:phasediagram}. First, note that a kink
(Fig.~\ref{fig:phasediagram-kink}) behaves quite differently from an antikink
(Fig.~\ref{fig:phasediagram-antikink}). For instance, a kink is never completely trapped
or reflected by an impurity. The reason is that it has zero intrinsic potential energy and
thus, no potential energy to lose during a scattering event. As a collective object, the
kink experiences a flat potential landscape along the chain. It will always go through the
impurity, unless $k_i$ is so soft or $v_0$ is so large that the initial kinetic energy of
the kink is sufficient to stretch the impurity spring to form a pinned antikink. That is
when scattering results in the kink being split. This also explains the positive slope of
the boundary line between these two regimes. (The topological constraints of the field
require that the number of kinks minus the number of antikinks remains
constant~\cite{Manton}, which is one for our boundary conditions.)

For an antikink, the scattering phase diagram has more regimes
(Fig.~\ref{fig:phasediagram-antikink}). The positive slope of the boundary curve at higher
$k_i$ between the upper reflection regime (square) and the transmission regime (circle)
comes from the fact that the higher the barrier is, the faster the antikink needs to be,
to get transmitted. The negative slope of the boundary between the transmission regime
(circle) and the trapping regime (triangle), suggests that a softer impurity spring causes
the antikink to dissipate more energy. The antikink then needs a sufficiently high initial
velocity to avoid being trapped at such an impurity site. The positive slope of the curve
between the trapped regime (triangle) and the lower reflection regime (square) suggests
that if the impurity spring is so soft such that it can no longer transform the kinetic
energy into other forms or channelize the kinetic energy to the other side of the impurity
sufficiently ``quickly'', an antikink incident with sufficiently high energy will then be
completely reflected. (In simulations we find that the maximum initial velocity with which
we can launch an antikink is around $v_0=12$. Above this, the antikink itself becomes
unstable and tends to quickly disintegrate.)

For the topological rotor chain, the antikink scattering behaviour is therefore very
similar to the ones reported for kinks and antikinks in previous studies on the $\phi^4$
model~\cite{Fei1992,Fraggis1989}. In addition, for normal $\phi^4$ kinks and antikinks,
one also observes resonance windows which are alternating regimes of the excitation being
reflected or trapped, along the axis of initial velocities for a given impurity strength.
These have not been observed during our simulations of the discrete topological chain.
Instead, we only observe a small range of alternating regimes where the antikink is
transmitted or trapped, around $k_i/k=0.75$ and $v_0=3.6$ in
Fig.~\ref{fig:phasediagram-antikink}. We leave a detailed characterization of the
resonance energy exchange between these modes for future studies.

\section{Effect of bond length impurities} \label{ch2}

In section IV we perform linear mode analysis of the topological chain, and in section VI we
study the nonlinear motion of (anti)kinks with impurities. Here in this section we will
show in a qualitative way that there is a connection between these two aspects. For
convenience, we investigate another type of impurity: the spring length.

\subsection{Linear mode analysis}

We start with a qualitative observation of the linear vibrational modes. For a perfect
topological rotor chain with free boundary conditions, there exists only one zero mode --
the translation mode of the kink. This is what the Maxwell-Calladine counting
predicts~\cite{Maxwell1864,Calladine1978}: the chain has $n$ rotors as degrees of freedom
and $n-1$ springs as constraints, and the former quantity minus the latter equals the
number of zero modes minus the number of states of self stress. (In a perfect chain there
is no states of self stress.) This counting does not depend on the geometrical parameters
of the chain components.

\begin{figure*}
  \subfloat[]{\label{fig:zero-mode}
   \includegraphics[width=0.9\textwidth]{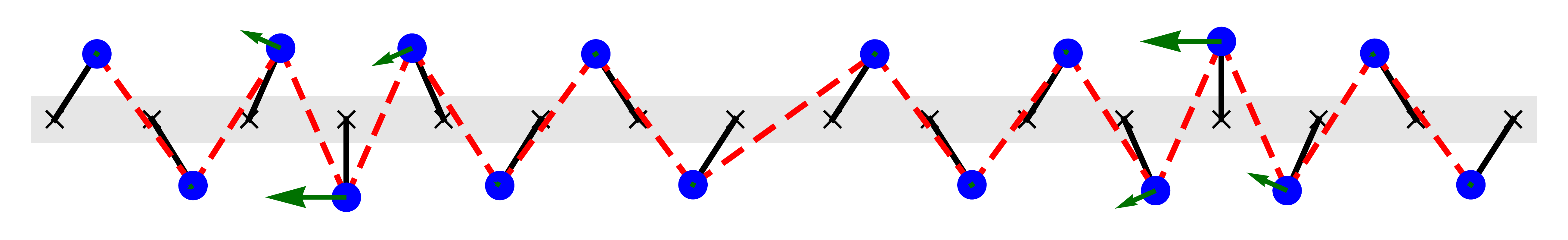}}
 
 \subfloat[]{\label{fig:soft-mode}
   \includegraphics[width=0.9\textwidth]{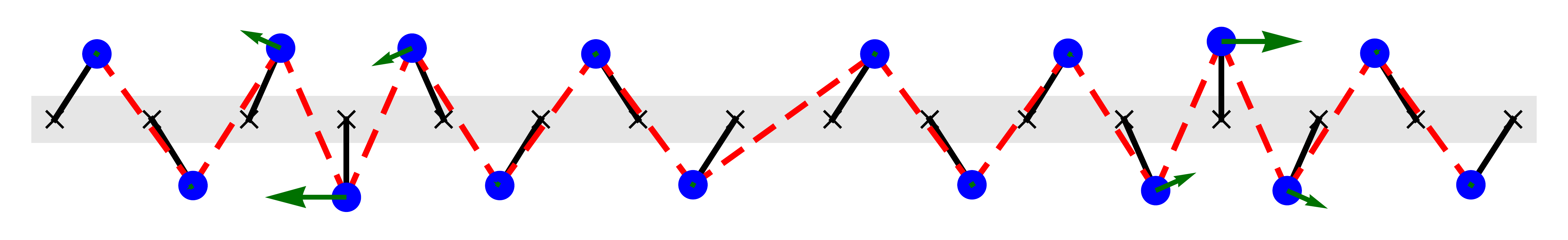}}
 
 \subfloat[]{\label{fig:tension-soft-mode}
   \includegraphics[width=0.9\textwidth]{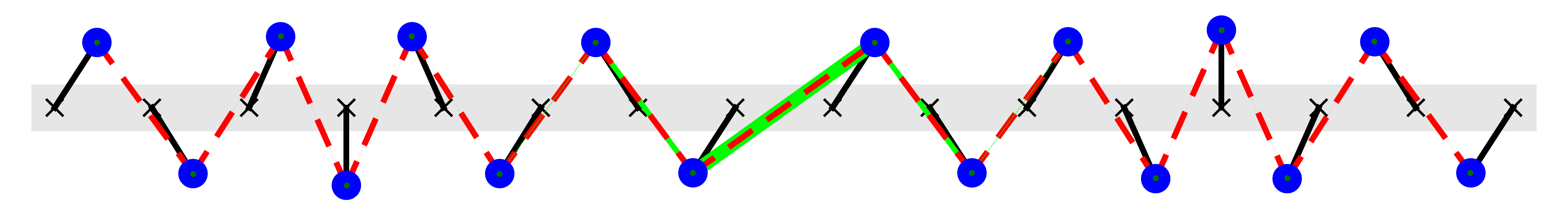}}
 
 \subfloat[]{\label{fig:lego-longbar-ak}
   \includegraphics[width=0.9\textwidth]{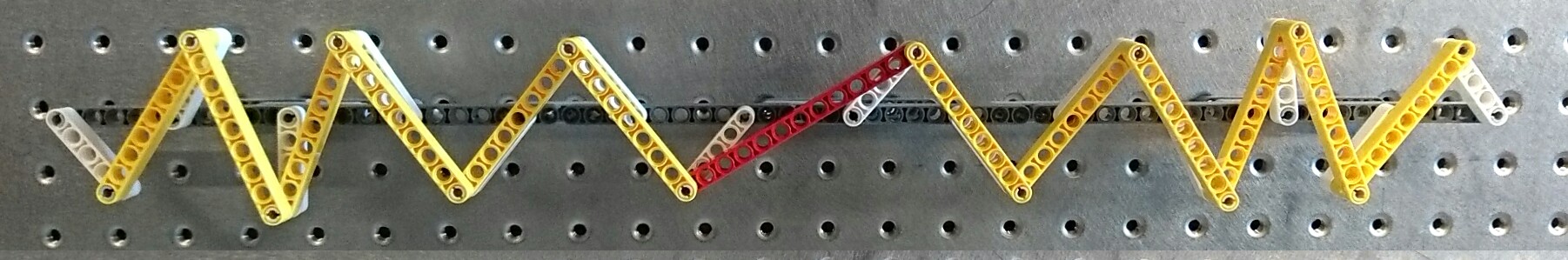}}
 
 \caption{\label{fig:longbar} The zero vibrational mode (\textbf{a}), the soft vibrational
   mode (\textbf{b}), and the soft tensional mode (\textbf{c}) of a topological chain with
   a longer spring in the middle as an impurity. The configuration parameters are
   $\overline{\theta}=0.58$, $r/a=0.8$, $\overline{l}/a=1.68$, $l_{0}/a=2.30$ and
   $l_{critical}/a=2.31$. The soft mode frequency is $7.7 \times 10^{-9}$ in the unit of
   $(r/a)\sqrt{k/M}$, which means the mode is much ``softer'' than the kink shape mode
   whose frequency is of the order $10^{-2}$. In (\textbf{a}) and (\textbf{b}), the arrows
   indicate the mode amplitude of the displacement of each rotor. In (\textbf{c}), the
   thickness of the green bars indicates the tensional mode amplitude on each spring. All
   the springs, both normal ones and the impurity, have the same stiffness. (\textbf{d})
   shows a LEGO demonstration. }
\end{figure*}

Now we increase one geometrical parameter, namely the length of the
middle spring $l_0$, so that it is an impurity in the system (Fig.~\ref{fig:longbar}). As long as no state of self stress is created, there
remains only one zero mode. However, as $l_0$ approaches a critical value $l_{critical}$,
several qualitative changes take place: (1) The profile of the chain varies significantly.
There are two kinks, one on each side of the impurity spring. (2) Eigenmode analysis shows
that the amplitude of the zero mode has two prominent parts that are spatially separated,
each of which is localized around a kink as an individual translation mode. Both parts of
the zero mode point towards the same direction. (3) An additional soft vibrational mode
appears, whose amplitude also has two separated parts just like the zero mode. But the
directions of these two parts are opposite to each other. This soft mode has a frequency
close to zero, much lower than that of kink shape modes. (4) A soft tensional mode dual to
the soft vibrational mode emerges, being localized around the impurity spring. (A
tensional mode is a vector whose components are the infinitesimal spring tensions caused
by the infinitesimal motion of the dual vibrational mode. The duality comes from the fact
that the tensional mode is an eigenfunction of the supersymmetrical ``partner'' of the
dynamical matrix, while the vibrational mode is an eigenfunction of just the dynamical
matrix. See~\cite{Kane2013a, Vitelli2014, Lubensky2015} for more details.)

% BGC:
% It may be worth defining ``tensional mode''. ``dual'' should
% probably also be explained.  

These changes do not contradict the Maxwell-Calladine counting: only one vibrational mode
has strictly zero frequency, unless $l_0$ actually reaches $l_{critical}$. In that case,
the frequencies of both the soft vibrational mode and the soft tensional mode go to zero.
By definition the tensional mode becomes a state of self stress. Then the
Maxwell-Calladine counting still holds as there are now two zero modes and one state of
self stress.

The above analysis only considers infinitesimal oscillations around
zero-energy equilibrium points. In the next section, we study qualitatively
the nonlinear motion of kinks with finite energy, providing a perspective complementary to
the linear analysis.

\subsection{Nonlinear dynamics: linkage limit}

\subsubsection{Setup: Hamiltonian}

To simplify the problem, we consider the linkage limit, where all the springs in a perfect
chain are non-deformable rigid bars so that they are holonomic constraints.% even at
%finite energy. 
There is only one degree of freedom which is the translational motion of
the kink. We choose the kink position $x$ as a collective variable to describe this degree
of freedom.

% BGC: I remove ``even at finite energy'' because that is a little
% confusing here; before we add the impurity, this system has no
% potential energy

\begin{figure*}
  \subfloat[]{\label{fig:defect-chain-conf}
                \includegraphics[width=0.9\textwidth]{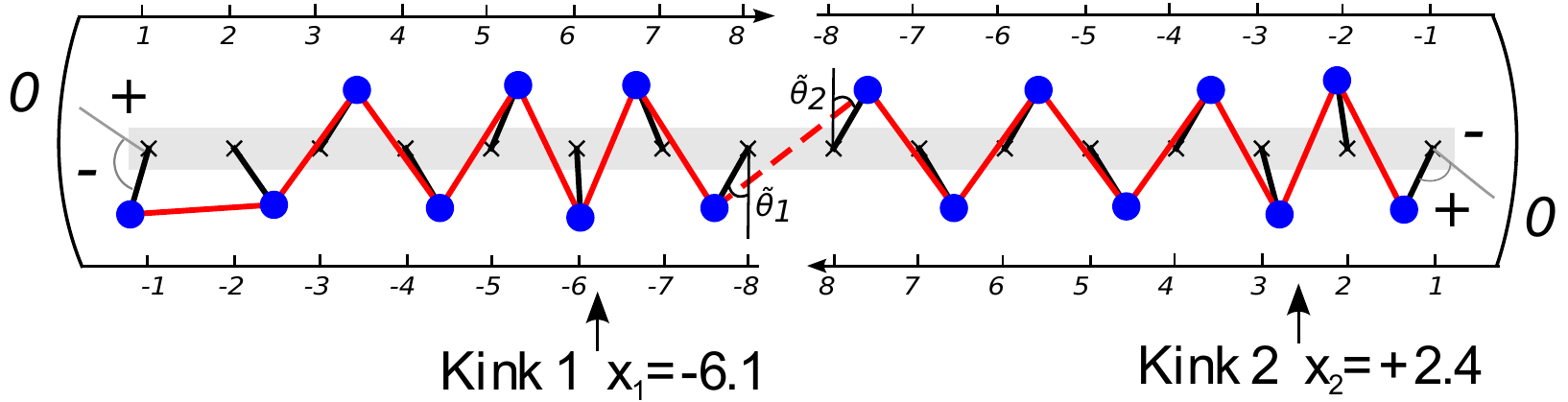}}
                
  \subfloat[]{\label{fig:KLphaseportrait}
                \includegraphics[width=0.9\textwidth]{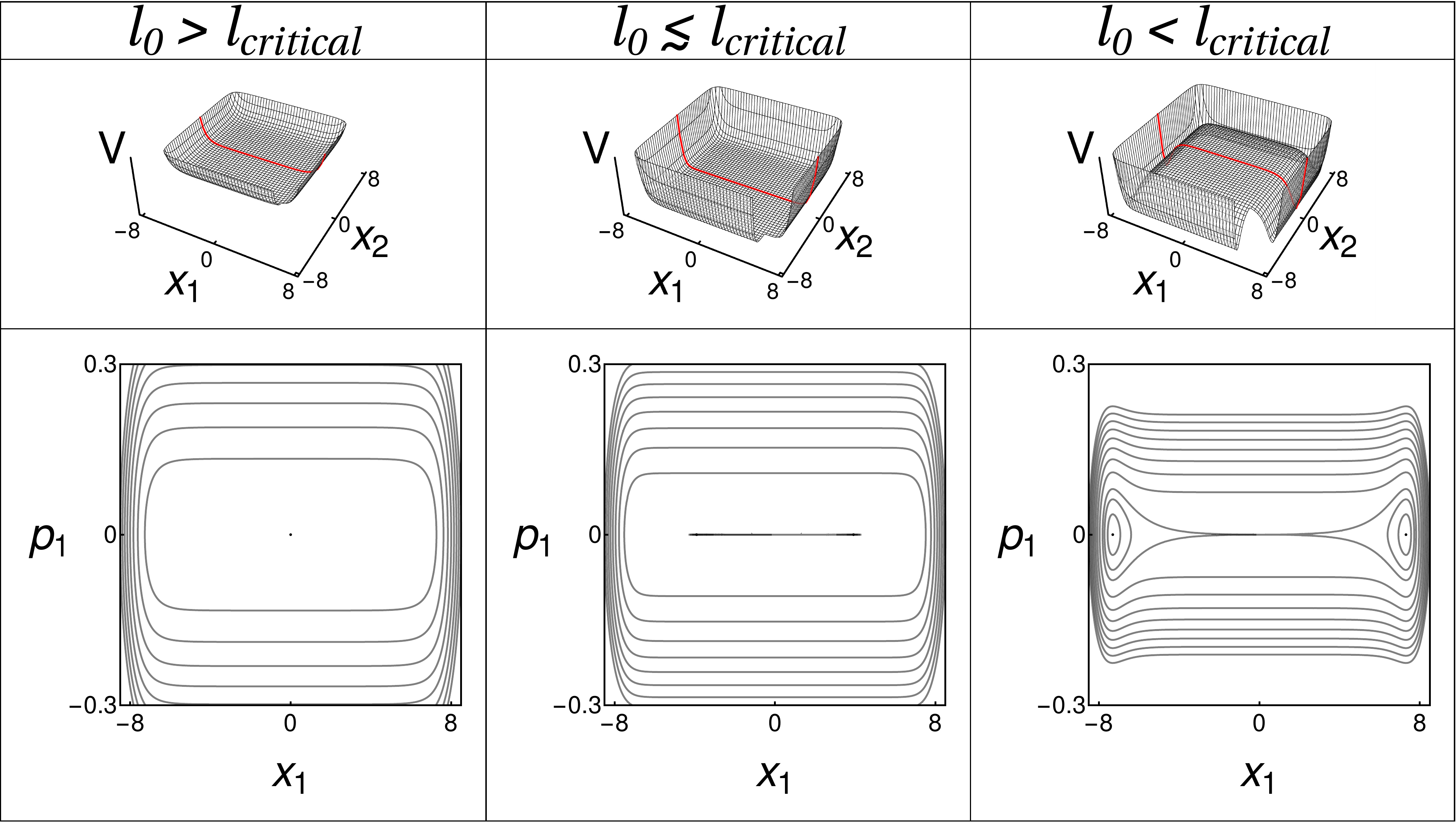}}
            
              \caption{\label{fig:defectiveKL} (\textbf{a}) Illustration of the coordinate
                system of a topological rotor linkage chain with $\overline{\theta}=0.58$,
                $r/a=0.8$, $\overline{l}/a=1.68$ and $l_{critical}/a=2.31$. The linkage
                bars are the solid lines and the impurity spring is the dashed line. In
                (\textbf{b}), the upper panels show the potential functions in 2D
                configuration space for various $l_0$. One corner of the function is
                trimmed for visualization. The red curve corresponds to the potential for
                Kink 1 in the one d.o.f.~case where Kink 2 is fixed at $x_2=0$. The lower
                panels show the phase portraits of Kink 1. }
\end{figure*}

Then we introduce the impurity by replacing the middle rigid bar with a longer 
spring that is ``soft'' (i.e.~with a finite spring constant) 
(Fig.~\ref{fig:defect-chain-conf}). A soft spring does not strictly constrain the
angles of the two rotors it connects but rather gives a potential energy to deviations
from its preferred length. The chain then has one fewer constraint, which in turn means
that it has two degrees of freedom. We regard the whole chain as two linkage sub-chains,
then the two degrees of freedom are shared by the two kinks of the sub-chains, which we
call Kink 1 and Kink 2 with position $x_1$ and $x_2$ respectively. The coordinate system
for the discrete chain model is illustrated in Fig.~\ref{fig:defect-chain-conf}, and its
precise definition is contained in Appendix \ref{App: AppendixC}. We see that by taking
the linkage limit, the number of degrees of freedom is reduced from the number of rotors
(16 for the chain in Fig.~\ref{fig:defect-chain-conf}) to the number of kinks (2 for two
kinks).

Now we derive the Hamiltonian. Note that the potential energy only comes from the
deformation of the impurity spring, which in turn just depends upon the angles of the head
rotors $\tilde{\theta}_{i}$. Since $x_i$ is the degree of freedom, it determines the state
of the sub-chain $i$, including $\tilde{\theta}_{i}$. Thus from the continuum theory
(Eq.~\ref{eq:kinksolution} where $u=r\sin\theta$), we obtain $\tilde{\theta}_i(x_i)$:
\begin{equation}
\sin \tilde{\theta}_i(x_i)= \sin\overline{\theta} \tanh \bigg(\frac{r
\sin\overline{\theta}(|x_i|-\tilde{x}_i)}{a^2} \bigg),
\end{equation}
where $\overline{\theta}$ is the equilibrium angle of a perfect chain, $a$ is the
lattice spacing, $r$ is the rotor length, and $\tilde{x}_i$ is the position of the head
rotor.

Putting $\tilde{\theta}_i(x_i)$ into the Hookean spring
potential $V=\frac{1}{2}k (l_{1,2}-l_0)^2$ where $l_{1,2}$ takes the form in
Eq.~(\ref{eq:instantlength}) and $l_0$ is the rest length of the impurity spring, we
obtain the potential function $V(x_1,x_2;l_0)$ as a function of the kink positions
(Fig.~\ref{fig:KLphaseportrait}). We formally define the effective kink momentum $p$ and
mass $m$ for the sub-chains in terms of the total kinetic energy of the rotors
$T=\sum_{j=1}^8 \frac{1}{2}m r^2 \dot{\theta}_j^2\equiv \frac{1}{2m}p^2$. Thus the
Hamiltonian $H(x_1,x_2,p_1,p_2;l_0)=T(p_1,p_2)+V(x_1,x_2;l_0)$ is obtained.

\subsubsection{Individual kink: Phase portrait}

We first investigate a simple case where Kink 2 is fixed at $x_2=0$ and only Kink 1 is
allowed to move. Then the chain has only one degree of freedom $x_1$. With the
Hamiltonian, we draw the phase portraits of $x_i$ for various $l_0$ in
Fig.~\ref{fig:KLphaseportrait}. We find that there is a critical value for the rest length
of the impurity spring
\begin{equation}
l_{critical}=\sqrt{(2r\sin\overline{\theta}+a)^2+(2r\cos\overline{\theta})^2},
\end{equation}
which determines the pattern of the phase portrait and the qualitative behavior of the 
dynamics of the chain.

% BGC: l_{critical} is used many times below so it's worth putting it
% on its own line

% The physical meaning of $l_{critical}$ will be explained later.

When $l_0<l_{critical}$, the dumbbell-shaped separatrix curve extends almost across
the whole reachable region of $x_1$. The two equilibrium points at $x_1 \approx +8$ and
$x_1 \approx -8$ correspond to the kink being localized around the impurity spring. $x_1$
is either positive or negative depending on the orientation of the end
rotor. At these two
equilibrium points the impurity spring is not stretched.

The behavior of Kink 1 depends on whether $E$ is above or below the separatrix curve's
energy $E_c=\frac{1}{2}k(l_0-l_{critical})^2$. If $E<E_c$, the trajectory in the phase
plane stays inside the region enclosed by separatrix and circulates around one of the
equilibrium points. In real space, Kink 1 makes small oscillations around the impurity
spring at either $x_1 \approx -8$ or $x_1 \approx +8$. If $E>E_c$ the trajectory moves in
the region outside of the separatrix. In real space, Kink 1 is able to go over the
sub-chain end and move back and forth between $x_1 \approx -8$ and $x_1 \approx +8$.

When $l_0$ approaches $l_{critical}$ from below and exceeds $l_{critical}$, the
separatrix curve shrinks and disappears. The two equilibrium points merge into one at
$x_1=0$ at the end of the sub-chain~\footnote{In the language of dynamical systems, this
process is called a \emph{supercritical pitchfork bifurcation}.}. In real space, the kink with
finite energy oscillates around the sub-chain end $x_1=0$.

\subsubsection{Two kinks: Accessible configuration space }

\begin{figure*}
  \subfloat[]{\label{fig:orbita}
    \includegraphics[width=0.45\textwidth]{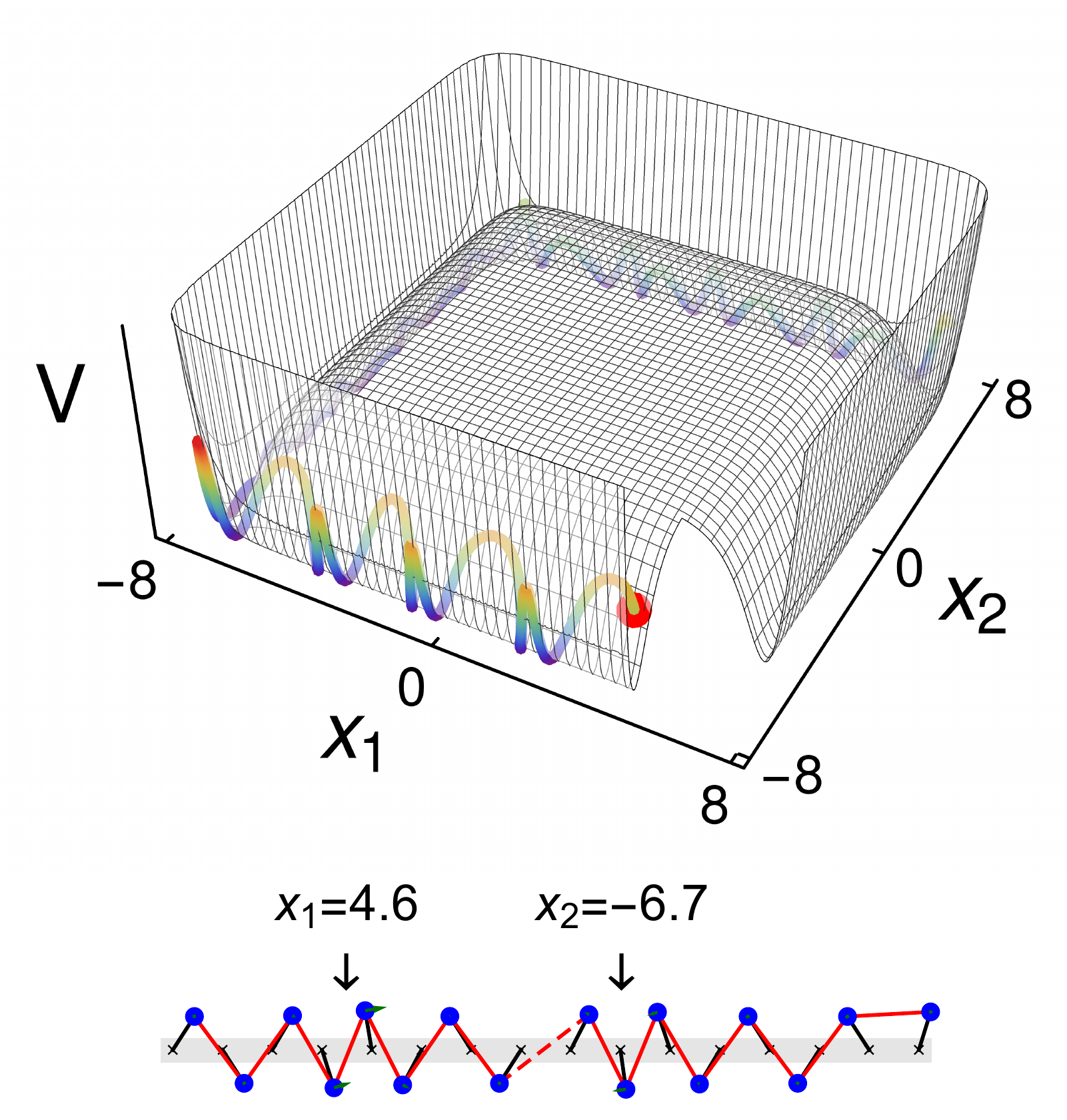}}
\subfloat[]{\label{fig:orbitb}
                \includegraphics[width=0.45\textwidth]{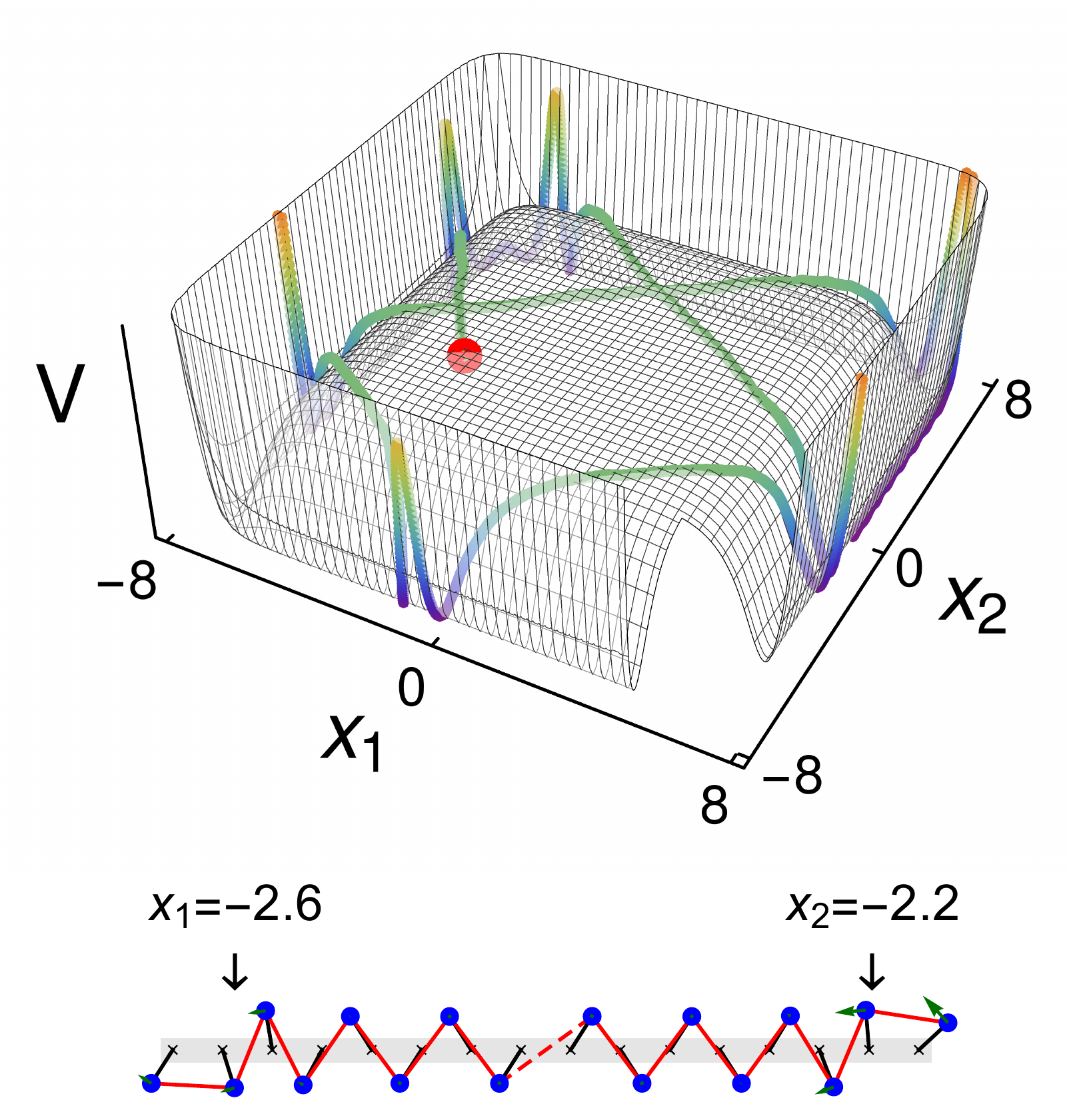}}
                
\subfloat[]{\label{fig:orbitc}
                \includegraphics[width=0.45\textwidth]{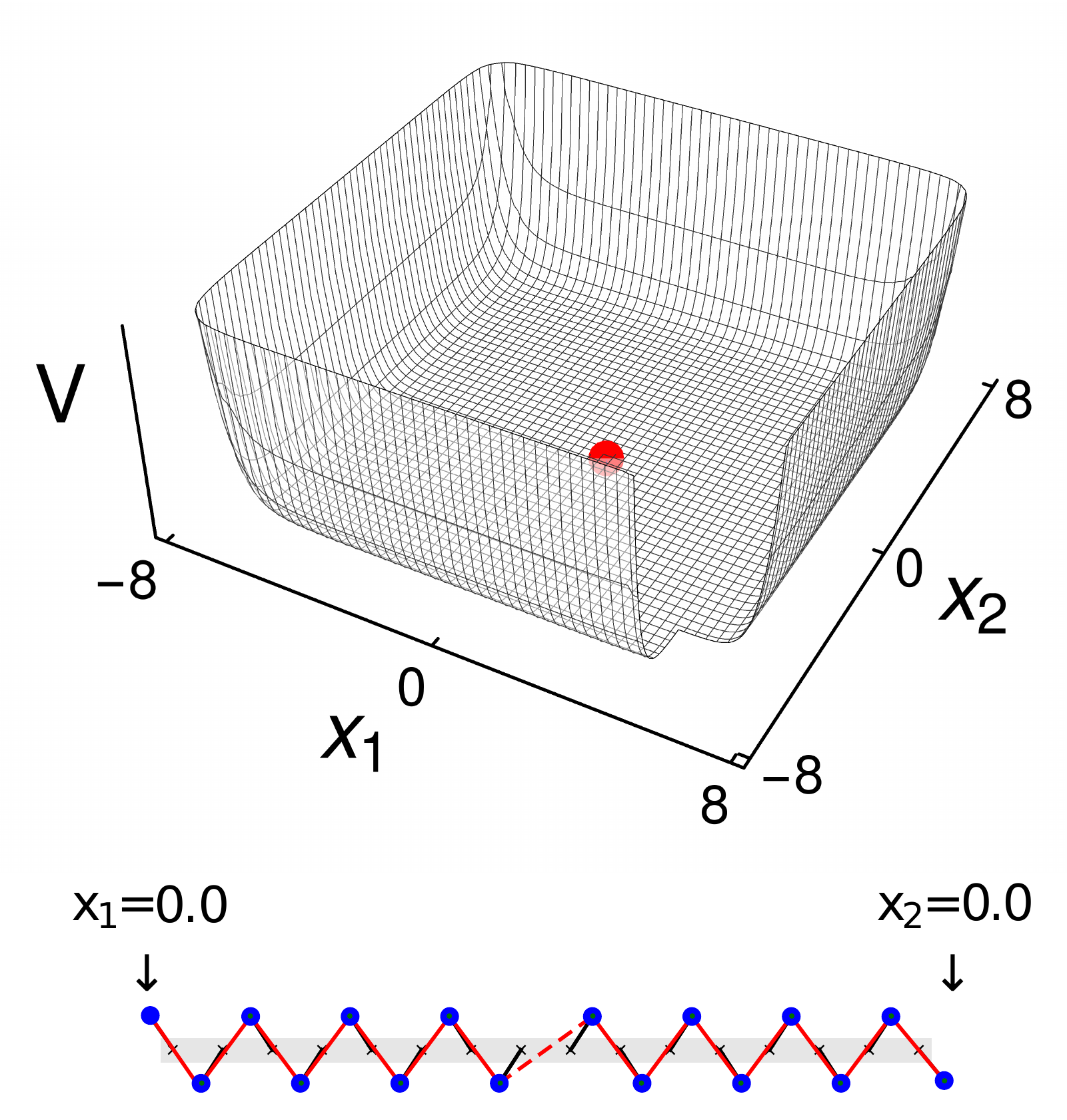}}
\subfloat[]{\label{fig:orbitd}
                \includegraphics[width=0.45\textwidth]{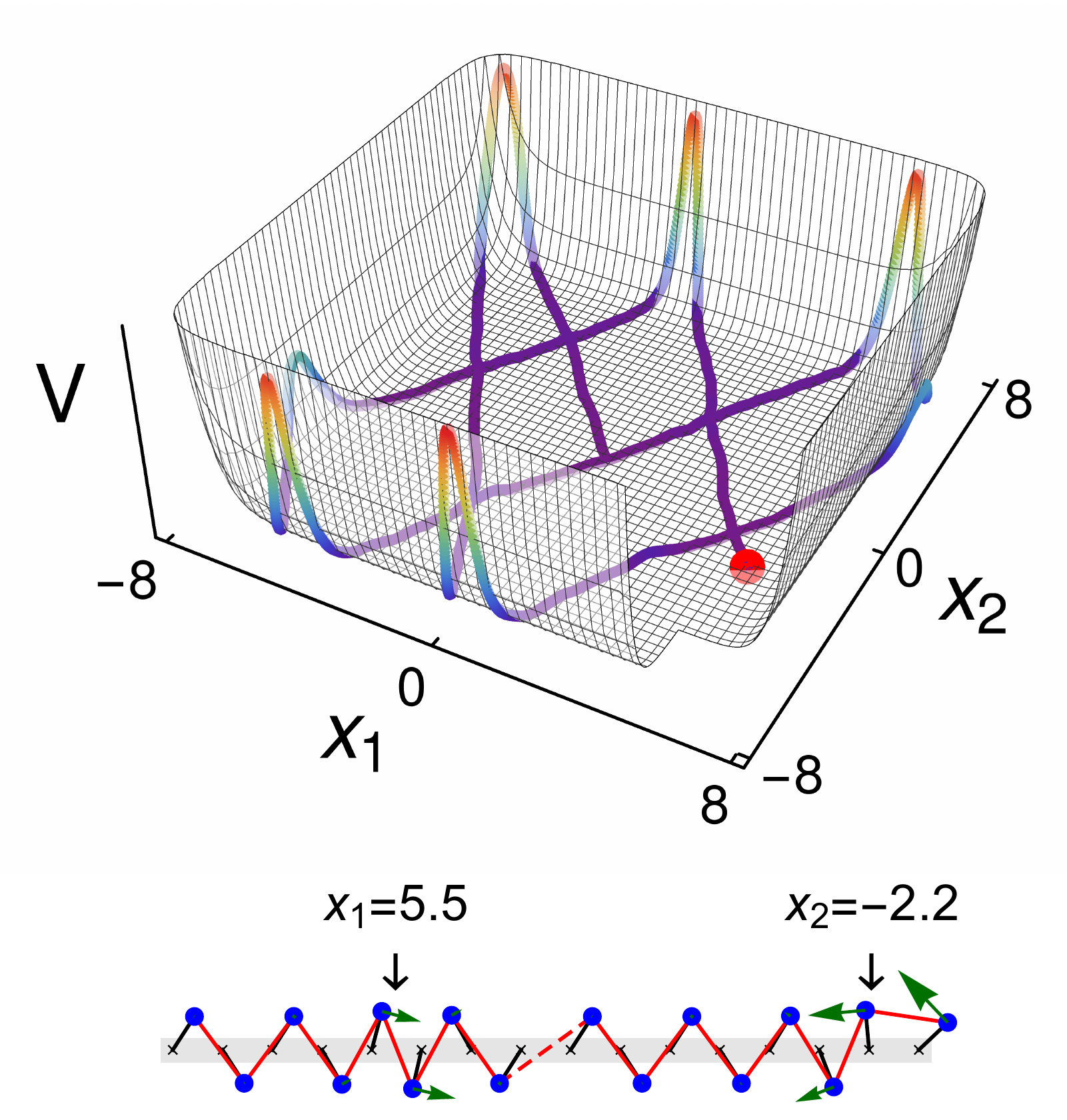}}
                  
              \caption{\label{fig:orbit}The trajectories of the chain generated by
                simulations of Newtonian dynamics on the theoretical potential function in
                the configuration space at \textbf{(a)} $l_0<l_{critical}$, $E<E_c$,
                \textbf{(b)} $l_0<l_{critical}$, $E>E_c$, \textbf{(c)} $l_0=l_{critical}$,
                $E=E_c=0$, and \textbf{(d)} $l_0>l_{critical}$. In the top figures of
                \textbf{(a)} and \textbf{(b)}, the color scale of the trajectories
                indicates the potential energy of the chain in arbitrary units. The big
                red dots correspond to the configuration of the real-space chains shown in
                the bottom figures of each panel.}

\end{figure*}

The phase space of a chain with two kinks is 4D. For the convenience of visualization, we
investigate the potential function $V(x_1,x_2;l_0)$ in the 2D configuration space. The
shape of the potential depends on $l_0$ and determines the qualitative dynamics of the two
kinks. We also perform simulations of Newtonian dynamics to investigate the qualitative
behavior of the nonlinear motion of the kinks.

When $l_0<l_{critical}$ (Fig.~\ref{fig:orbita}), the potential looks like a square
Mexican hat. The bottom of potential valley is a square ring, on which all the points are
at zero energy. In linear mode analysis, we find a zero mode along the valley and a
soft mode along the transverse direction. We will show that the nonlinear dynamics at
finite energy possesses the traits that are closely related to those in the linear
analysis at zero energy.

Note that the impurity spring is maximally stretched at $x_1=x_2=0$, and the corresponding
potential maximum $E_c=\frac{1}{2}k(l_0-l_{critical})^2$. It is the minimal energy for
both kinks to move away from the impurity. If $E<E_c$, the two kinks
take turns moving on
their respective sub-chains. One kink oscillates near the impurity spring, while the other
kink moves away. The nonlinear dynamics of the kinks is visualized as a trajectory going
along the bottom of the potential valley. The accessible region in the configuration space
is a square annulus, at the corner of which the major part of energy is transferred from
the one kink to another.  In fact, this can be interpreted as the
motion of a single ``split'' kink through the system.

% BGC: I added the last sentence above -- does it make sense? That's
% how I thought about it ...

% Yujie: Yes. That is my thought, too.

When $E \geq E_c$ (Fig.~\ref{fig:orbitb}), there is sufficient energy for both kinks to
move away from the impurity spring simultaneously. In the configuration space, the
trajectory gets out of the potential valley and climbs up to the 2D plateau in the middle.
The accessible region now is a square disk. In real space, the kinks independently hit the
impurity spring and get reflected.

When $l_0=l_{critical}$ (Fig.~\ref{fig:orbitc}), the linear mode analysis predicts that
the chain model in Fig.~\ref{fig:orbitc} has two zero modes, each being localized around
the kink at the end of the respective sub-chain, and a state of self stress localized
around the impurity spring. From the viewpoint of nonlinear dynamics, the potential
function changes qualitatively: As $l_0$ approaches $l_{critical}$, the square ring of the
potential valley shrinks into one point at $x_1=x_2=0$, and $E_c$ goes to zero. In other
words, the Mexican hat transforms into a single basin. In this shrinking process, the soft mode,
which corresponds to the oscillation transverse to the valley, transitions into a zero
mode, because the depth of the valley vanishes. In terms of nonlinear dynamics, this
transition means that no matter how small the total energy $E$ is, the accessible region
in the configuration space is always a square disk rather than a square annulus.

% BGC: I added "single basin"; hope that is accurate

% Yujie: I put "single basin" to the previous sentence, replacing the "oval cup". And the
% accessible region is about the 2D space, while basin seems to describe the potential
% landscape.

When $l_0>l_{critical}$ (Fig.~\ref{fig:orbitd}), the impurity spring is compressed,
which gives a minimum potential energy $E_{min}=\frac{1}{2}k(l_0-l_{critical})^2$ for the
static configuration. In a linear analysis, the two zero modes become normal modes with
finite frequency, as the impurity spring pushes the two kinks to the chain ends,
generating a finite restoring force for the motion of the modes. In
the nonlinear dynamics, the accessible region of the kinks is still a square disk.

\begin{figure}[h!] \centering
  \includegraphics[width=0.45\textwidth]{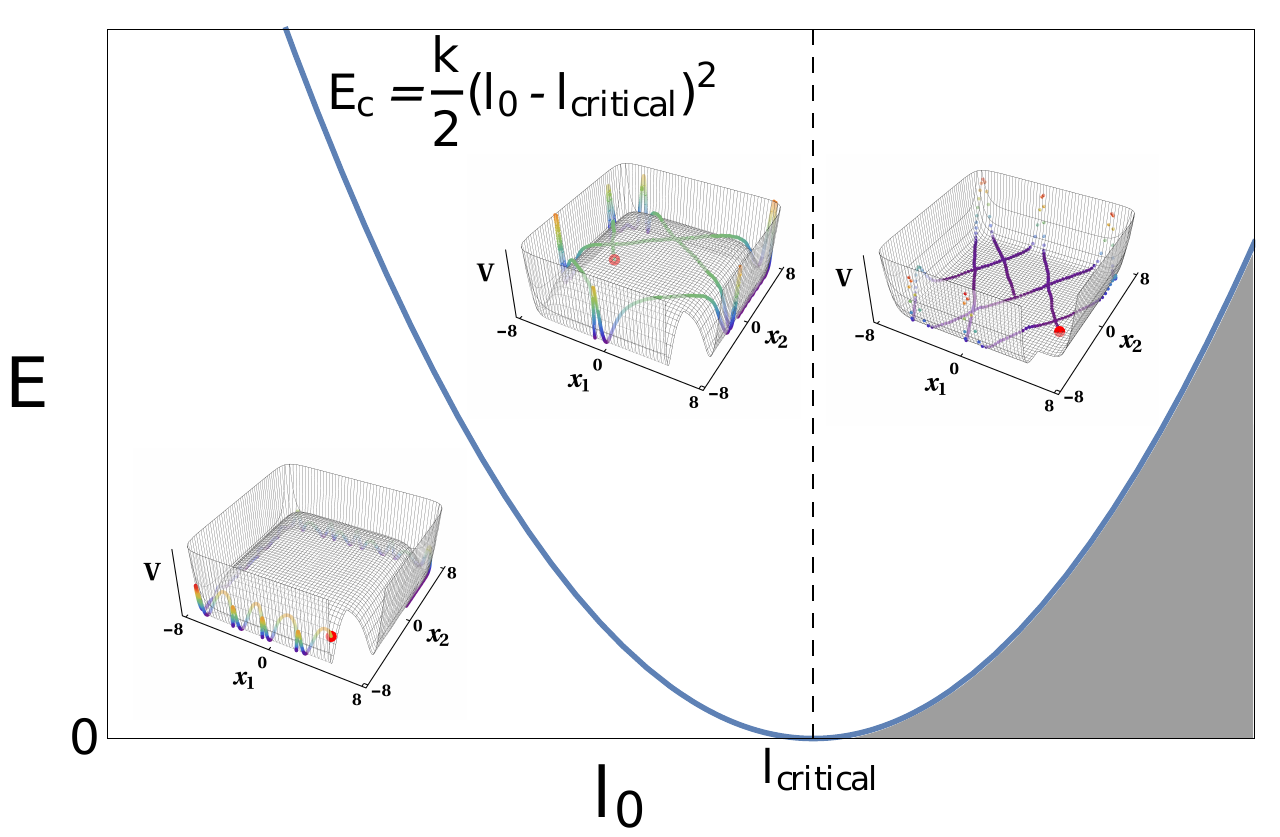}
  \caption{The parameter space of the total energy $E$ and the impurity spring length
$l_0$. The critical energy $E_c$ as a function of $l_0$ forms a parabola. The chain shows
different dynamical behaviors across the left branch of the parabola. The vertical dashed
line of $l_0 = l_{critical}$ is the boundary line across which the shape of the potential
function transitions qualitatively. The gray area below the right branch of the parabola
is energetically forbidden.}
  \label{fig:ELdiagram}
\end{figure}

Fig.~\ref{fig:ELdiagram} summarizes the above results with $E$ and $l_0$ as parameters.
When $l_0 \leq l_{critical}$, the curve $E_c=\frac{1}{2}k(l_0-l_{critical})^2$ marks the
transition of the accessible region in configuration space from an annulus to a disk. Note
that we only investigate the case of $l_0>\overline{l}$, in which Fig.~\ref{fig:ELdiagram}
is valid. For $l_0<\overline{l}$ case, the potential landscape takes a different form, and
so does the possible transition. We do not cover this case in this paper, however, as we
have made the connection between linear mode analysis and nonlinear dynamics.

% % For real chains with finite energy away from the linkage limit, all the springs, whether
% % the impurity one or the others, are no longer constraints. In principle the description
% of % the full nonlinear dynamics of the chain should include all degrees of freedom of
% the % rotors. However, at the low energy limit, the translational motion of the kinks
% captures % the essential dynamics of the real chain in 2D configuration space formed by
% the two kinks % still provides a good approximation .

\section{Conclusion}

We have studied the nonlinear dynamics of a topological rotor chain.
% BGC: I think we can omit the rest of this sentence
% which is a
%diatomic chain of rotors coupled with springs, whose unit cell breaks spatial inversion
%symmetry. 
The continuum limit is well-approximated by a modified $\phi^4 $ theory whose
nonlinear excitations are the kinks and antikinks. We have seen how the breaking of
inversion asymmetry at the discrete level results in an asymmetry between the kink and
antikink excitations that affects the properties of linear modes around these excitations,
their transport along an ordered lattice as well as in how these excitations interact with
a lattice impurity. The results herein further enrich the class of phenomenon described by
the $\phi^4$ theory, a model which is extensively studied and finds numerous applications
in many fields of physics.

Some questions for further research -- (1) We find that kinks reflect perfectly off the free
boundaries of a topological rotor chain. This is surprising given that in the continuum
limit, the $\phi^4$ kink is a non-integrable solution and thus could create bound states
or emit radiation as it interacts with a free boundary. Furthermore, an antikink cannot
reach a free boundary without colliding with a kink -- another feature which we do not yet
know how to interpret within the continuum theory. (2) We have not undertaken a detailed
study of the phases of motion for an antikink (wobbling, spinner). Preliminary simulations
indicate that antikink configurations in these other phases are in fact unstable. The large 
amount of initial
spring stretching energy necessary in a configuration where the rotors point ``away from
each other'' is immediately converted to kinetic energy and induces rapid spinning of the
nearby rotors which then spreads across the system in a chaotic fashion. It is not clear
how this effect would arise in the continuum theory, which for the spinner, is related to
the integrable sine-Gordon model~\cite{Chen2014}.

A more speculative question is whether there are connections between our results and the
observed asymmetry between kinks and antikinks in certain one-dimensional quantum magnetic
systems, called delta or sawtooth chains~\cite{Nakamura1996,Sen1996}. These systems also
have two uniform ground states which may be thought of as the analog of ``right-leaning''
and ``left-leaning'' states, and also share the property that the excitation energy for a
kink is zero while for an antikink is large and finite.

\begin{acknowledgments}

This work was supported by FOM and NWO. We thank J. Paulose and A. Souslov for fruitful
conversations and critical reading of the manuscript.

\end{acknowledgments}

\appendix
\section{Complex notation}\label{App: AppendixA}

We use complex variables to derive the explicit relation between neighbouring rotor
angles. Adopting the notation in Fig.~\ref{fig:two-rotor}, we put the pivot of rotor 1 at
the origin of complex plane and the pivot of rotor 2 at the coordinate~(a,0). The
positions of the rotor tips are
\begin{align} z_1 &= ire^{-i\theta_1}, \\ z_2 &= a-ire^{i\theta_2}.
\end{align}

We have two constraints (where a bar represents complex conjugations):
\begin{align} (z_2-z_1)(\bar{z}_2-\bar{z}_1)=l^2_0, \\ (z_2-a)(\bar{z}_2-a)=r^2.
\end{align}

Eliminating $\bar{z}_2$ from above two constraints, we find a quadratic equation for
$z_2$,
\begin{align} Az^2_2 + Bz_2 + C = 0,
\end{align} where
\begin{align} A &= \frac{\bar{z}_1-a}{a-z_1}, \\ B &=
\left(\frac{l^2_0+a^2-2r^2}{a-z_1}\right)-a\left(\frac{\bar{z}_1-z_1}{a-z_1}\right), \\ C
&= a^2-r^2-a\left(\frac{l^2_0+a^2-2r^2}{a-z_1}\right).
\end{align}

We have two branches of the solution for $z_2$
\begin{align} z_2= \frac{-B \pm \sqrt{B^2-4AC}}{2A},
\end{align} which explicitly expresses the black curve in Fig.~\ref{fig:kcobwebplot}.

\section{Vibrational modes of prestressed mechanical structures: Method of Tangent
  stiffness matrix}
\label{App: AppendixD}

\begin{figure}[h!]
  \subfloat[]{\label{fig:prestress-config1}
    \includegraphics[width=0.22\textwidth]{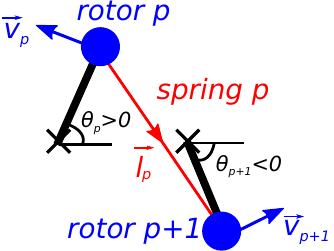}}
  \subfloat[]{\label{fig:prestress-config2}
  \includegraphics[width=0.22\textwidth]{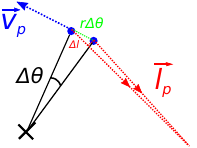}}
  \caption{\label{fig:prestress-config}Detailed configurations around a single spring $p$.}
\end{figure}

Consider a single spring $p$ in the configuration shown in Fig.~\ref{fig:prestress-config1} (note here, we are now specifying rotor angles $\theta$ with respect to the positive $x$-axis). From geometry, we find
\begin{align}
  \label{eq:prestress1}
  \begin{split}
    f_p &= -\vec{v}_p \cdot \vec{l}_p~~\hat{t}_p\\
    f_{p+1}&= \vec{v}_{p+1} \cdot \vec{l}_p~~\hat{t}_p.
  \end{split}                                        
\end{align}
Here, $f_p$ is the spring force projected along the tangent vector $\vec{v}_p$ of
rotor $p$
\begin{align}
  \begin{split}
    \vec{v}_p=
    \begin{pmatrix}
      -\sin\theta_p\\
      \cos\theta_p
    \end{pmatrix}.
  \end{split}                                        
\end{align}
$\vec{l}_p$ is the vector along the length of the spring $p$ and points from rotor $p$ to $p+1$,
\begin{align}
  \begin{split}
    \vec{l}_p=
    \begin{pmatrix}
      a+r\cos\theta_{p+1}-r\cos\theta_p\\
      r\sin\theta_{p+1}-r\sin\theta_p
    \end{pmatrix}.
  \end{split}                                        
\end{align}
$\hat{t}_p$ is a scalar \textit{tension coefficient} for spring $p$, defined as $\hat{t}_p\equiv t_p/|\vec{l}_p|$, where $t_p\equiv k_p(|\vec{l}_p|-\overline{l})$ for a harmonic spring. Here, $|\vec{l}_p|$ is the instantaneous length of spring $p$, $\overline{l}$ is the rest length of the spring, and $k$ is the spring constant.

In order to find the tangent stiffness, we differentiate Eqn.~(\ref{eq:prestress1}) with respect to the rotor angles $\theta_p$ and $\theta_{p+1}$
\begin{align}
  \label{eq:prestress2-1}
  \begin{split}
\frac{\partial f_p}{r\partial \theta_p}&=\frac{\partial( -\vec{v}_p \cdot \vec{l}_p)
}{r\partial \theta_p}~~\hat{t}_p-\vec{v}_p \cdot \vec{l}_p~\frac{\partial \hat{t}_p}{r\partial \theta_p}
  \end{split}                                        
\end{align}
\begin{align}
  \label{eq:prestress2-2}
  \begin{split}
\frac{\partial f_p}{r\partial \theta_{p+1}}&=\frac{\partial (-\vec{v}_p \cdot \vec{l}_p)
p}{r\partial \theta_{p+1}}~~\hat{t}_p-\vec{v}_p \cdot \vec{l}_p~\frac{\partial \hat{t}_p}{r\partial \theta_{p+1}}
  \end{split}                                        
\end{align}
\begin{align}
  \label{eq:prestress2-3}
  \begin{split}
\frac{\partial f_{p+1}}{r\partial \theta_p}&=\frac{\partial( \vec{v}_{p+1} \cdot \vec{l}_p)
}{r\partial \theta_p}~~\hat{t}_p+\vec{v}_{p+1} \cdot \vec{l}_p~\frac{\partial \hat{t}_p}{r\partial \theta_p}
  \end{split}                                        
\end{align}
\begin{align}
  \label{eq:prestress2-4}
  \begin{split}
\frac{\partial f_{p+1}}{r\partial \theta_{p+1}}&=\frac{\partial( \vec{v}_{p+1} \cdot \vec{l}_p)
}{r\partial \theta_{p+1}}~~\hat{t}_p+\vec{v}_{p+1} \cdot \vec{l}_p~\frac{\partial \hat{t}_p}{r\partial \theta_{p+1}}.
  \end{split}                                        
\end{align}

To simplify Eqn.~(\ref{eq:prestress2-1}), we express
\begin{align}
    \label{eq:prestress3-1}
  \begin{split}
    \frac{\partial \hat{t}_p}{r\partial \theta_p}&=\frac{\d \hat{t}_p}{\d |\vec{l}_p|}~\frac{\partial
      |\vec{l}_p|}{r\partial\theta_p}
  \end{split}                                        
\end{align}
\begin{align}
    \label{eq:prestress3-2}
  \begin{split}
    \frac{\d \hat{t}_p}{\d |\vec{l}_p|}&=\frac{\d (t_p/|\vec{l}_p|)}{\d |\vec{l}_p|}=\frac{1}{|\vec{l}_p|}(g_p-\hat{t}_p)=\hat{g}_p/|\vec{l}_p|,
  \end{split}                                        
\end{align}
where $g_p\equiv \d t_p/d|\vec{l}_p|$ is defined as the axial stiffness and
$\hat{g}_p\equiv g_p-\hat{t}_p$ is defined as the \textit{modified axial stiffness}.

From Fig.~\ref{fig:prestress-config2}, we see that $\Delta l=r\Delta \theta ~ (-\vec{v}_p \cdot \vec{l}_p)/|\vec{l}_p|$ and therefore,
\begin{align}
    \label{eq:prestress4}
  \begin{split}
    \frac{ \partial |\vec{l}_p|}{r\partial\theta_p}=\frac{(-\vec{v}_p \cdot \vec{l}_p)}{|\vec{l}_p|} 
\end{split}                                        
\end{align}

Substituting Eqn.~(\ref{eq:prestress3-1} - \ref{eq:prestress4}) into Eqn.~(\ref{eq:prestress2-1}), we find
\begin{align}
    \label{eq:prestress5a}
  \begin{split}
    \frac{\partial f_p}{r\partial \theta_p}&=\frac{\partial( -\vec{v}_p \cdot
      \vec{l}_p)}{r\partial \theta_p}~~\hat{t}_p
    -(\vec{v}_p \cdot \vec{l}_p)~\frac{\hat{g}_p}{|\vec{l}_p|} \frac{(-\vec{v}_p \cdot \vec{l}_p)}{|\vec{l}_p|} .
\end{split}                                        
\end{align}
Similarly, we simplify Eqns.~(\ref{eq:prestress2-2} - \ref{eq:prestress2-4}) 

With the above derivatives, we can now define the tangent stiffness matrix. For a single spring $p$, the tangent stiffness matrix, $\mathbf{K}_p$, relates small changes in rotor position to small changes in rotor forces
\begin{align}
    \label{eq:prestress5b}
  \begin{split}
    \begin{pmatrix}
      \delta f_p\\
      \delta f_{p+1}
    \end{pmatrix}
    = \mathbf{K}_p
    \begin{pmatrix}
      r\delta\theta_p\\
      r\delta\theta_{p+1}
    \end{pmatrix}
  \end{split}                                        
\end{align}
and can be expressed as
\begin{align}
    \label{eq:prestress6}
  \begin{split}
 \mathbf{K}_p=
    \begin{pmatrix}
      n_p\\
      n_{p+1}
    \end{pmatrix}
    \Big[\hat{g}_p\Big]
    \begin{pmatrix}
      n_p &n_{p+1}
    \end{pmatrix}
    +\mathbf{s_p},
  \end{split}                                        
\end{align}
where $n_p\equiv -\vec{v}_p \cdot \vec{l}_p/|\vec{l}_p|$, $n_{p+1}\equiv -\vec{v}_{p+1} \cdot \vec{l}_p/|\vec{l}_p|$ and the \textit{stress matrix}
$\mathbf{s}_p$ is
\begin{align}
    \label{eq:prestress7}
  \begin{split}
    \mathbf{s}_p=
    \begin{pmatrix}
      -\frac{\partial(\vec{v}_p \cdot\vec{l}_p)}{r\partial \theta_p}~\hat{t}_p &~~
      -\frac{\partial(\vec{v}_p \cdot \vec{l}_p)}{r\partial \theta_{p+1}}~\hat{t}_p\\
      \\
      \frac{\partial(\vec{v}_{p+1} \cdot \vec{l}_p)}{r\partial \theta_p}~\hat{t}_p &~~
      \frac{\partial(\vec{v}_{p+1} \cdot \vec{l}_p)}{r\partial \theta_{p+1}}~\hat{t}_p 
    \end{pmatrix}.
  \end{split}                                        
\end{align}

To derive the total tangent stiffness $\mathbf{K}$ for the rotor chain,  we first represent the tangent stiffness $\mathbf{K}_p$ in a global coordinate system as an $n \times n$ matrix,
and then sum up all the $\mathbf{K}_p$ for the $n-1$ springs:
\begin{align}
    \label{eq:prestress8}
  \begin{split}
    \mathbf{K}=\sum_{p=1}^{n-1}\mathbf{K}_p=\sum_{p=1}^{n-1}\mathbf{a}_p\big[\hat{g}_p\big]\mathbf{a}_p^T+\sum_{p=1}^{n-1}\mathbf{S}_p,
  \end{split}                                        
\end{align}
where 
\begin{align}
    \label{eq:prestress9a}
  \begin{split}
    \mathbf{a}_p=
    \begin{pmatrix}
      0\\
      \vdots\\
      0\\
      n_p\\
      n_{p+1}\\
      0\\
      \vdots\\
      0
    \end{pmatrix}
  \end{split}                                        
\end{align}
and
\begin{align}
    \label{eq:prestress9b}
  \begin{split}
    \mathbf{S}_p=
    \begin{pmatrix}
       \text{\huge 0} & ~~~~~~\ldots&& \text{\huge 0}\\
       \vdots &\mathbf{s}_{p11}~&\mathbf{s}_{p12}~& \vdots\\
      \vdots&\mathbf{s}_{p11}~&\mathbf{s}_{p12}~&\vdots\\
      \text{\huge 0} & ~~~~~~\ldots&& \text{\huge 0}
    \end{pmatrix}.
  \end{split}                                        
\end{align}
In $\mathbf{a}_p$, the $n_p$ and $n_{p+1}$ terms are in the $p$th and $p+1$th row respectively, and all the other terms are zero. In $\mathbf{S}_p$, $\mathbf{s}_{pij}$ is
the $(i,j)$ element of the $2\times2$ stress matrix $\mathbf{s}_p$ for a single spring $p$ and is located in the $(p-1+i,p-1+j)$ position of $\mathbf{S}_p$, and all the other terms in
$\mathbf{S}_p$ are zero. Here $\mathbf{S}_p$ has a simpler form than that of Ref.~\cite{Guest2006} because we exploit the fact that only nearest neighbours are coupled
in the topological chain.

\section{Simulation methods}\label{App: AppendixB}

The molecular dynamics simulations are carried out in Mathematica. The ODEs are solved by
the function NDSolve, which uses a multi-step method (LSODA) by default.

In the simulations, we set the lattice spacing $a=1$, the rotor mass $M=1$, and an
arbitrary time unit $t=1$. The spring constant $k$ is measured in units of $M/t^2$. The
linear velocity of a rotor is measured in units of $a/t$. The initial velocity $v_0$ of a
(anti)kink is defined as the velocity amplitude of the unit translation mode ${\bf e}^t$
and $e^t_i$ is the mode component on the $i$-th rotor. Thus the initial kinetic energy is
$\Sigma_i \frac{1}{2} m (v_0 e^t_i)^2 = \frac{1}{2} m v_0^2$.

\section{Peierls-Nabarro potential barrier via continuum theory}
\label{app:AppendixPN}

We derive the PN potential by discretizing the potential energy density in the continuum
theory, i.e.~taking the quasi-continuum limit. The PN potential is, by definition, the
potential that the kink faces as it propagates along the adiabatic trajectory (ad.~tr.) :
\begin{align}
  \label{eq:PNdef}
  \begin{split}
   V_{PN}(X)&=V(...,u_{n-1},u_n,u_{n+1},...)|_{X\in ad.tr.}.
  \end{split}                                        
\end{align}
Here, $X$ is the position of the (anti)kink center, $u_n$ is the continuum field at lattice site $n$, $V$ is a discretization of the potential energy density $V(\theta)$ in Eqn.~(\ref{eq:continuouspotential}) and is obtained by summing the potential $f(n,X)$ of each lattice site:
\begin{align}
  \label{eq:PNsum}
  \begin{split}
V(...,u_{n-1},u_n,u_{n+1},...)&=\sum f(n,X),
  \end{split}                                        
\end{align}
where
\begin{align}
  \label{eq:PNoneterm1}
  \begin{split}
    f(n,X)&=\frac{2k}{\overline{l}^2}\left(\frac{a^2}{2}\frac{\d u_n}{\d (na)}+\overline{u}^2-u^2_n\right)^2.
  \end{split}                                        
\end{align}
$f(n,X)$ is the approximate potential at a single site $n$ when the (anti)kink center is at
$X$. Here, we discretize the continuum potential energy density rather than directly use
the exact form of the lattice potential in Eqn.~(\ref{eq:discretepotential2}), so that
we can readily substitute $u_n$, the continuum field at site $n$, into $f(n,X)$ which
results in an integrable solution. We choose the static solution ($v=0$) of
Eqn.~(\ref{eq:kinksolution}) as the adiabatic trajectory:
\begin{align}
  \label{eq:PNstaticsolution}
  \begin{split}
    u_n(X)=\pm\overline{u}\tanh\Big(\frac{na-X}{w}\Big),
  \end{split}                                        
\end{align}
where the ``$+$'' is for the antikink, ``$-$'' is for the kink, and the width of the (anti)kink $w=\frac{a^2}{2r\sin\overline{\theta}}$~\cite{Chen2014}. Substituting Eqn.~(\ref{eq:PNstaticsolution}) into Eqn.~(\ref{eq:PNoneterm1}), we find
\begin{align}
  \label{eq:PNoneterm2}
  \begin{split}
    f(n,X)&= 0~~~~~~~~~~~~~~~~~~~~~~~~~~~~~\textrm{for the kink,}\\
    f(n,X)&=\frac{8k\overline{u}^4}{\overline{l}^2}\sech^4\Big(\frac{na-X}{w}\Big)~~\textrm{for the antikink.}
  \end{split}                                        
\end{align}

Thus $V_{PN}(X)=0$ for the kink, in accordance with the fact that the kink configuration does not stretch springs and hence costs zero potential energy. For the antikink, we use the Poisson summation formula to express:
\begin{align}
  \label{eq:PNpoisson}
  \begin{split}
    V_{PN}(X)&=\sum_{n=-\infty}^{+\infty} f(n,X)=\sum_{k=-\infty}^{+\infty} \hat{f}(k,X)\\
         &=\sum_{k=-\infty}^{+\infty}\int_{-\infty}^{+\infty}\d nf(n,X)e^{-2\pi
i k n}.
  \end{split}
\end{align}
To leading order, we only consider the first harmonic terms $k=1$ and
$k=-1$ ($k=0$ recovers the continuum approximation). For $k=1$, we find
\begin{align}
  \label{eq:PNB1}
  \begin{split}
    &~~\int_{-\infty}^{+\infty}\d nf(n,X)e^{-2\pi i n}\\
    % &=\int_{-\infty}^{+\infty}\d n\frac{8k\overline{u}^4}{\overline{l}^2}\sech^4\Big(\frac{na-X}{w}\Big) e^{-2\pi i n}\\
    % &=\int_{-\infty}^{+\infty}\d (n-X/a)\frac{8k\overline{u}^4}{\overline{l}^2}\sech^4\Big(\frac{(n-X/a)a}{w}\Big)\\
    % &~~~~~~~~~~~~~~~~~~~~~~~~~~~~~~~~~~~~~e^{-2\pi i(n-X/a)}e^{-2\pi i(X/a)}\\
    &=e^{-2\pi i(X/a)} \int_{-\infty}^{+\infty}\d n'\frac{8k\overline{u}^4}{\overline{l}^2}\sech^4\Big(\frac{n'a}{w}\Big)
    e^{-2\pi in'}.
    % &=\frac{1}{2}V_{PNB}e^{-2\pi i(X/a)}
  \end{split}
\end{align}

The complex exponential suggests a sinusoidally varying potential along the coordinate $X$ of the adiabatic trajectory, with a period that is equal to the lattice spacing $a$. We
define the \textit{PN barrier} ($V_{PNB}$) as the height of this sinusoidal potential. The last integral in Eqn.~(\ref{eq:PNB1}) can be completed using residues to yield
\begin{align}
  \label{eq:PNBappendix}
  \begin{split}
    V_{PNB}% &=\int_{-\infty}^{+\infty}\d n'\frac{16k\overline{u}^4}{\overline{l}^2}\sech^4\Big(\frac{n'a}{w}\Big)
    % e^{-2\pi in'}\\
    &=\frac{4\pi^2\big(\pi^2+(a/w)^2\big)}{3\big(1+4(r/a)^2-(a/w)^2\big)\sinh(\pi^2w/a)}\\
    &\propto e^{-\pi^2w/a}~~~~~~~~~~~~\textrm{for large $w/a$.}
  \end{split}
\end{align}

\section{Definition of kink coordinates in discrete models }\label{App: AppendixC}

The concept of kinks stems from the continuum $\phi^4$ theory. To extend this concept to
the discrete chain model, we define the coordinate system of a sub-chain kink as follows
(Fig.~\ref{fig:defect-chain-conf}): The absolute value of the position of a kink equals
the rotor's integer index if the rotor is vertical, otherwise the position is a real
number interpolating between the indices of the two neighboring rotors that are leaning
opposite to each other. The positional interpolation is proportional to the linear
interpolation between the absolute values of the angles of two neighbor rotors. The rotor
angles are the measured against the vertical alternatively, as mentioned in Section
\ref{sec:cobweb}. When a kink approaches the end points of the chain, the end rotor flips
over. Here the kink profile from the continuum theory ceases to be
valid. Thus we take as our convention that a kink is at the origin of
the coordinate system when the end rotor is collinear with the
spring connecting to the next rotor, and its sign depends on whether the end rotor leans
upwards or downwards. The coordinate between $0$ and $1$ (or $-2$) is obtained by linear
interpolation of the angles of the end rotor at $0$ and $1$ (or $-2$). In this \textit{ad
 hoc} convention, the chain forms a state of self stress when both kinks are at origin.
The two sub-chains are aligned head-to-head, and the two head rotors ($|x_i|=8$) are
coupled by the impurity spring.

%BGC: I replaced ``definition'' by ``convention''.

\clearpage\end{CJK*}

\bibliography{library}% Produces the bibliography via BibTeX.

\end{document}